\documentstyle[12pt,epsf]{article}

\def\abs#1{\left| #1\right|}
\def\sgn{\mathop{\rm sgn}}
\def\etmiss{\slashchar{E}_T}
\def\fb{{\rm fb}}

\def\Meff{M_{\rm eff}}
\def\Msusy{M_{\rm SUSY}}
\def\lsp{\tilde\chi_1^0}
\def\ra{\rightarrow}
\def\GeV{{\rm GeV}}

\def\mhalf{m_{1/2}}
\def\tchi{\tilde\chi}
\def\tg{\tilde g}
\def\tq{\tilde q}

\let\badcite=\cite
\def\cite{~\badcite}

\def\slashchar#1{\setbox0=\hbox{$#1$}           
   \dimen0=\wd0                                 
   \setbox1=\hbox{/} \dimen1=\wd1               
   \ifdim\dimen0>\dimen1                        
      \rlap{\hbox to \dimen0{\hfil/\hfil}}      
      #1                                        
   \else                                        
      \rlap{\hbox to \dimen1{\hfil$#1$\hfil}}   
      /                                         
   \fi}                                         %

\def\dofig#1#2{\centerline{\epsfxsize=#1\epsfbox{#2}}}

\begin{document}
\begin{titlepage}
\begin{flushright}
LBNL-39412
\end{flushright}
\bigskip\bigskip

\centerline{\Large\bf Precision SUSY Measurements at LHC\footnotemark}

\footnotetext{This work was supported in part by the Director, Office
of Energy Research, Office of High Energy and Nuclear Physics,
Division of High Energy Physics of the U.S. Department of Energy under
Contracts DE-AC03-76SF0009 and DE-AC02-76CH00016 and by the Swedish National Research Council.}

\bigskip
\centerline{\bf I. Hinchliffe$^a$, F.E. Paige$^b$, M.D. Shapiro$^a$,
J. S\"oderqvist$^c$, and W. Yao$^a$}
\centerline{$^a${\it Lawrence Berkeley National Laboratory, Berkeley, CA}}
\centerline{$^b${\it Brookhaven National Laboratory, Upton, NY}}
\centerline{$^c${\it Royal Institute of Technology (KTH), Stockholm, Sweden}}
\bigskip

\begin{abstract}
	If supersymmetry exists at the electroweak scale, then it should be
discovered at the LHC. Determining masses, of supersymmetric particles however, is more
difficult.  In this paper,  methods are discussed to determine combinations
of masses and of branching ratios precisely from experimentally
observable distributions. In many cases such measurements alone can
greatly constrain the particular supersymmetric model and determine its parameters with an 
accuracy of a few percent. Most of the results shown correspond to
one year of running at LHC at ``low luminosity'', $10^{33}\,
{\rm cm^{-2}s^{-1}}$.
\end{abstract}
\end{titlepage}

\tableofcontents

\newpage
\section{Introduction}
\label{sec:intro}

	If supersymmetry (SUSY) exists at the electroweak scale, then
gluinos and squarks will be copiously produced in pairs at the LHC and
will decay via cascades involving other SUSY particles to the lightest
SUSY particle (LSP), the $\lsp$. In most models the $\lsp$ is stable,
must be neutral and therefore escapes the detector. It should then be
easy to observe deviations from the Standard Model such as an excess
of events with multiple jets plus missing energy $\etmiss$ or with
like-sign dileptons $\ell^\pm\ell^\pm$ plus
$\etmiss$\cite{ATLAS,CMS,BCPT}. Determining SUSY masses is more
difficult because each SUSY event contains two LSP's, and there are
not enough kinematic constraints to determine the momenta of these.

	The strategy developed in Ref.\cite{LHCsnow} and in this paper
involves three steps.  First, we use a simple inclusive analysis to
establish a deviation from the Standard Model. We select events with
at least four jets and large missing energy and plot the distribution
of
$$
\Meff = p_{T,1} + p_{T,2} + p_{T,3} + p_{T,4} + \etmiss\,.
$$
Typically, this is dominated by Standard Model processes at low
$\Meff$ but is a factor of 5--10 larger than the Standard Model
prediction for large $\Meff$. The value of $\Meff$ at which the signal
exceeds that standard model backgrounds provides a first estimate of the SUSY masses. 

	The second step in the strategy is to identify characteristic
signatures of particles occurring near the end of the SUSY decay cascades
and to use these as the starting point for further analysis. This is
best explained by an example. Suppose that gluinos are slightly
heavier than squarks and that $\tchi_2^0 \ra \lsp h$ is kinematically
allowed. Then one can have the following decay chain:
$$
\arraycolsep=0pt 
\begin{array}{llll}
\tg &&+\ & \tg \\
\downarrow &&& \downarrow \\
\tq_L \,+\; &\bar q && \tq_R + \bar q \\
\downarrow &&& \downarrow \\
\tchi_2^0 \,+\; &q && \lsp + q \\
\downarrow &&&  \\
\lsp \,+\; &h && \\
 &\downarrow &&\\
 & b + \bar b &&
\end{array}
$$
Such an event typically contains two hard jets from the $\tilde
q_{L,R}$ decays, two $b$ jets from the $h$ decay, large $\etmiss$, and
soft jets from the gluino decays and from gluon radiation. In this
case, one can reconstruct $h \ra b \bar b$ as a peak in the $b \bar b$
mass distribution and measure its mass\cite{ERW}. Then the $h$ can be
combined with either of the two hard jets. The smaller of these two
masses must be less than the squark mass and so has a sharp edge that
measures a known function of the $\tilde q_L$, $\tchi_2^0$, and
$\tchi_1^0$ masses. In many cases several such measurements can be
made to determine several combinations of masses more or less
precisely.

	Given actual data, the third step would be to make a global
fit of a SUSY model to all available measurements, including both the
precision measurements just described and more inclusive ones such as
the jet, lepton, and $b$-jet multiplicities and $p_T$ distributions.
Such an analysis involves simulating large numbers of signal samples
and is beyond the scope of this study. Instead, we try to determine
the SUSY parameters using just the precision measurements of
combinations of masses. In some cases this almost completely
determines the SUSY model, while in others it significantly constrains
it.

\begin{table}[t]
\caption{SUGRA parameters for the five LHC points.\label{params}} 
\begin{center}
\begin{tabular}{cccccc}
\hline\hline
Point & $m_0$ & $m_{1/2}$ & $A_0$ & $\tan\beta$ & $\sgn{\mu}$ \\
      & (GeV) & (GeV)   & (GeV)   &             &             \\
\hline
1 & 400 & 400 &   0 & \phantom{0}2.0 & $+$\\
2 & 400 & 400 &   0 & 10.0 & $+$\\
3 & 200 & 100 &   0 & \phantom{0}2.0 & $-$\\
4 & 800 & 200 &   0 & 10.0 & $+$ \\
5 & 100 & 300 & 300 & \phantom{0}2.1 & $+$\\
\hline\hline
\end{tabular}
\end{center}
\end{table}

	What precision measurements can be made is very dependent on
the SUSY model and so must be studied for specific SUSY parameters.
The ATLAS and CMS Collaborations at the LHC have been considering five
points in the minimal supergravity (SUGRA) model listed in
Table~\ref{params}\cite{LHCsnow}. The SUGRA model \cite{SUGRA} has the
minimal SUSY particle content; universal scalar masses $m_0$, gaugino
masses $\mhalf$, and trilinear breaking terms $A_0$ at the grand
unification scale; and radiative electroweak symmetry breaking driven by the large top
quark mass.  After electroweak breaking, the remaining parameters are
a ratio of vacuum expectation values $\tan\beta$ at the weak scale and
a sign, $\sgn\mu=\pm1$.  We assume a default value for the top quark
mass of 175~GeV and comment on the sensitivity to it below. While
this model may not be the one that nature has chosen, we would like to
emphasize that simulations can only be performed in the context of a
consistent model.  This is because many promising signals that might
be clearly distinguished from Standard Model backgrounds in one channel, can be
obscured by production and decays of other supersymmetric particles.
We may not believe in this model, but the model that nature has chosen
will be self-consistent.

	In the SUGRA model, if $\tchi_2^0 \ra \lsp h$ is kinematically
allowed, it has a substantial branching ratio and provides one
starting point. This is the case for LHC Points~1, 2, and 5. If this
decay is kinematically forbidden, then in many cases $\tchi_2^0 \ra
\lsp \ell^+\ell^-$ can be observed. This is the case for LHC Points~3
and 4.  The endpoint of the $\ell^+\ell^-$ mass distribution provides
a direct measure of $M(\tchi_2^0) - M(\lsp)$, and opposite-sign,
same-flavor dileptons can be used to identify events containing
$M(\tchi_2^0)$. Other modes exploited in this paper include $\tchi_2^0
\ra \tilde\ell_R^\pm \ell^\mp \ra \lsp \ell^\pm \ell^\mp$ and
$\tchi_4^0 \ra \tchi_1^\pm W^\mp \ra \lsp e^\pm \mu^\mp \nu\bar\nu$.

	All the analyses presented here are based on
isajet~7.22\cite{ISAJET} and a toy detector simulation. At least 50k
events were generated for each signal point. The standard model
background samples contained 250k events for each of $t \bar t$, $WZ$
with $W \to e\nu,\mu\nu,\tau\nu$, and $Zj$ with $Z \to
\nu\bar\nu,\tau\tau$, and 5000K QCD jets (including $g$, $u$, $d$,
$s$, $c$, and $b$) divided among five bins covering $50 < p_T <
2400\,\GeV$. The calorimeter energy resolutions are taken to be
\begin{eqnarray}
{\rm EMCAL} &\quad& 10\%/\sqrt{E} + 1\% \nonumber\\
{\rm HCAL}  &\quad& 50\%/\sqrt{E} + 3\% \nonumber\\
{\rm FCAL}  &\quad& 100\%/\sqrt{E} + 7\%,\ |\eta| > 3\,.\nonumber
\end{eqnarray} 
A uniform segmentation $\Delta\eta = \Delta\phi = 0.1$ is used with no
transverse shower spreading. This is particularly unrealistic for the
forward calorimeter. Jets are found using GETJET\cite{ISAJET} with a
fixed cone size $R=0.4$ or 0.7.  Missing transverse energy is
calculated by taking the magnitude of the vector sum of the transverse
energy deposited in in the calorimeter cells.  The jet multiplicity in
SUSY events is rather large, so we will use a cone size of 0.4 unless
otherwise stated.  A lepton efficiency of 90\% and a $b$-tagging
efficiency of 60\% is assumed.  Isolated leptons are required to satisfy an
isolation requirement that no more than 10 GeV of additional $E_T$ be
present in a cone of radius $R = 0.2$ around the lepton.   Results are presented for an integrated
luminosity of $10\,\fb^{-1}$, corresponding to one year of running at
$10^{33}\,{\rm cm^{-2}s^{-1}}$ so pileup has
not been included. We will occasionally comment on the
cases where the full luminosity of the LHC, {\it i.e.}\ $10^{34}\,{\rm
cm^{-2}s^{-1}}$, will be needed to complete the studies. For many of
the histograms shown, a single event can give rise to more than one
entry due to different possible combinations. When this occurs, all
combinations are included.

	In Section~\ref{sec:meff} of this paper, we discuss using
$\Meff$ to get a rough estimate of SUSY masses. We then turn to more
detailed analyses. In these, we shall usually make cuts so that the
Standard Model backgrounds are very small. These cuts are not optimal,
particularly in the case of the higher mass points where event rates
are lower. It may be desirable to have more signal events at the cost
of more background. In Sections~\ref{sec:lhc3} and \ref{sec:lhc4} we
discuss LHC Points~3 and 4 respectively. These points have rather
light gauginos, so $\tchi_2^0 \ra \lsp \ell^+\ell^-$ can be used to
measure $M(\tchi_2^0)-M(\lsp)$.  In Section~\ref{sec:lhc5} we discuss
LHC Point~5, which has both $\tchi_2^0 \ra \lsp h \ra \lsp b \bar b$
and $\tchi_2^0 \ra \tilde\ell^\pm \ell^\mp \ra \lsp \ell^\pm
\ell^\mp$. In Section~\ref{sec:lhc1} we briefly discuss LHC Points~1
and 2. These have gluino and squark masses of about 1~TeV, so they
really need more than $10\,\fb^{-1}$ for detailed study. In
Section~\ref{sec:scan} we investigate how well the precision
measurements discussed determine the parameters of SUSY model.
Finally, we draw some conclusions.

\newpage
\section{Effective Mass Analysis}
\label{sec:meff}

\begin{figure}[h]
\dofig{3.20in}{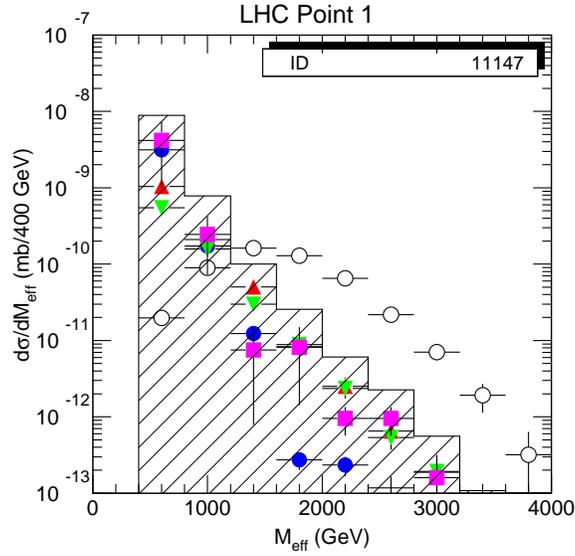}
\caption{LHC Point~1 signal and Standard Model backgrounds. Open
circles: SUSY signal.  Solid circles: $t\bar t$. Triangles: $W\ra\ell\nu$,
$\tau\nu$.  Downward triangles:  $Z\ra\nu\bar\nu$, $\tau\tau$.
Squares: QCD jets.  Histogram: sum of all
backgrounds.\label{point1_147}}
\end{figure}

\begin{figure}[h]
\dofig{3.20in}{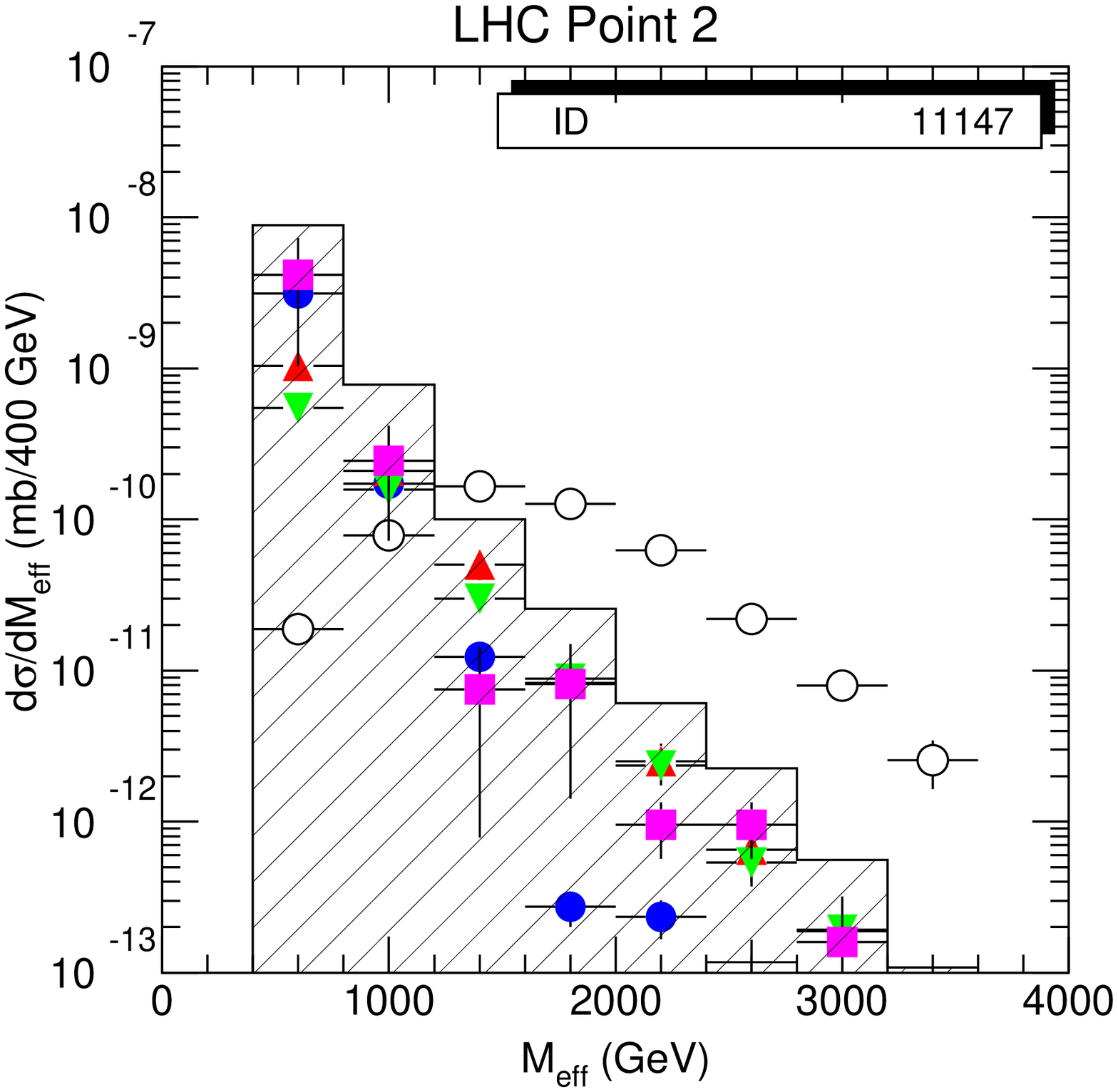}
\caption{SUSY signal and Standard Model backgrounds for LHC Point~2. See
Figure~1 for symbols.\label{point2_147}}
\end{figure}

\begin{figure}[h]
\dofig{3.20in}{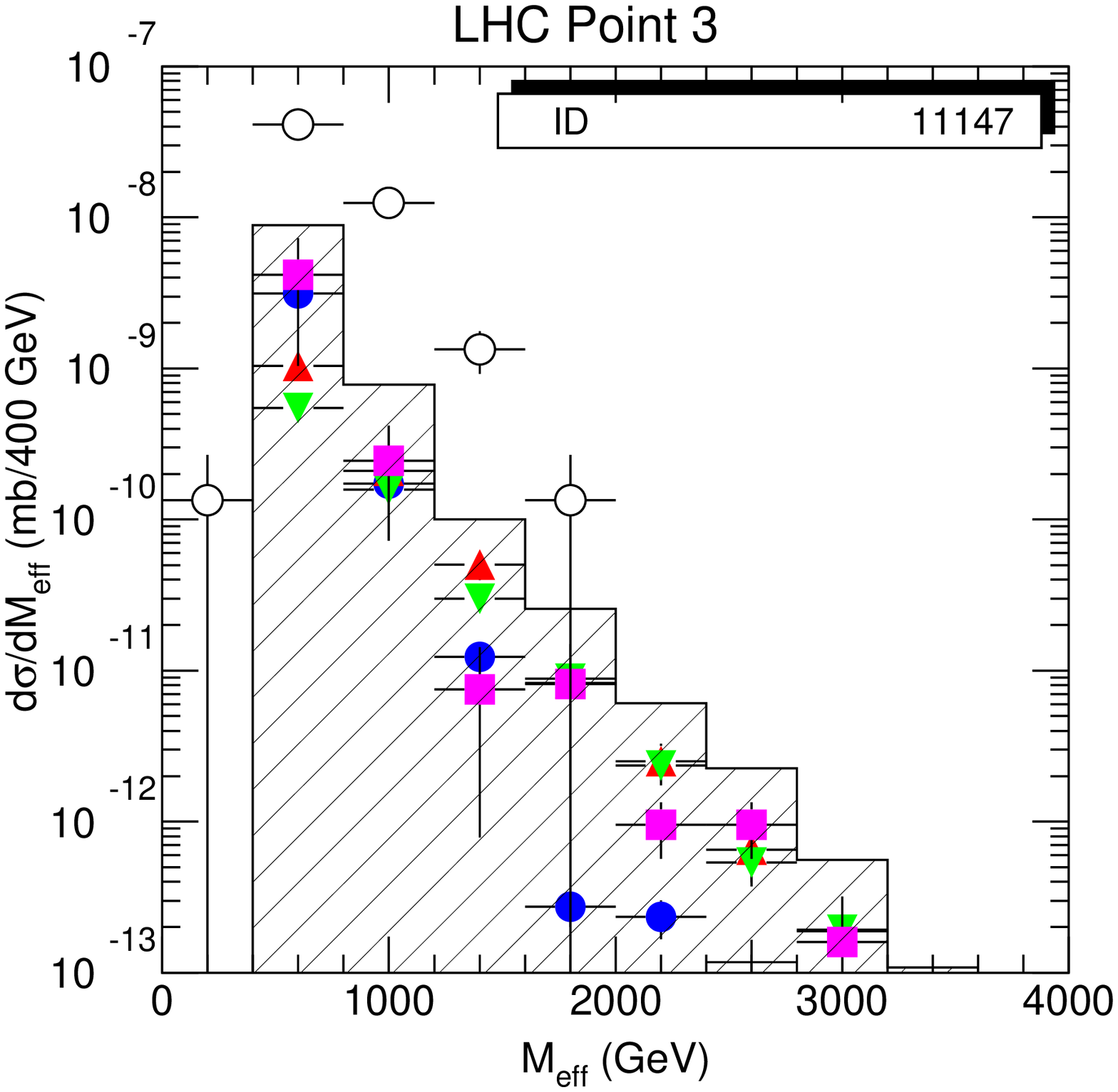}
\caption{SUSY signal and Standard Model backgrounds for LHC Point~3. See
Figure~1 for symbols.\label{point3_147}} 
\end{figure}

\begin{figure}[h]
\dofig{3.20in}{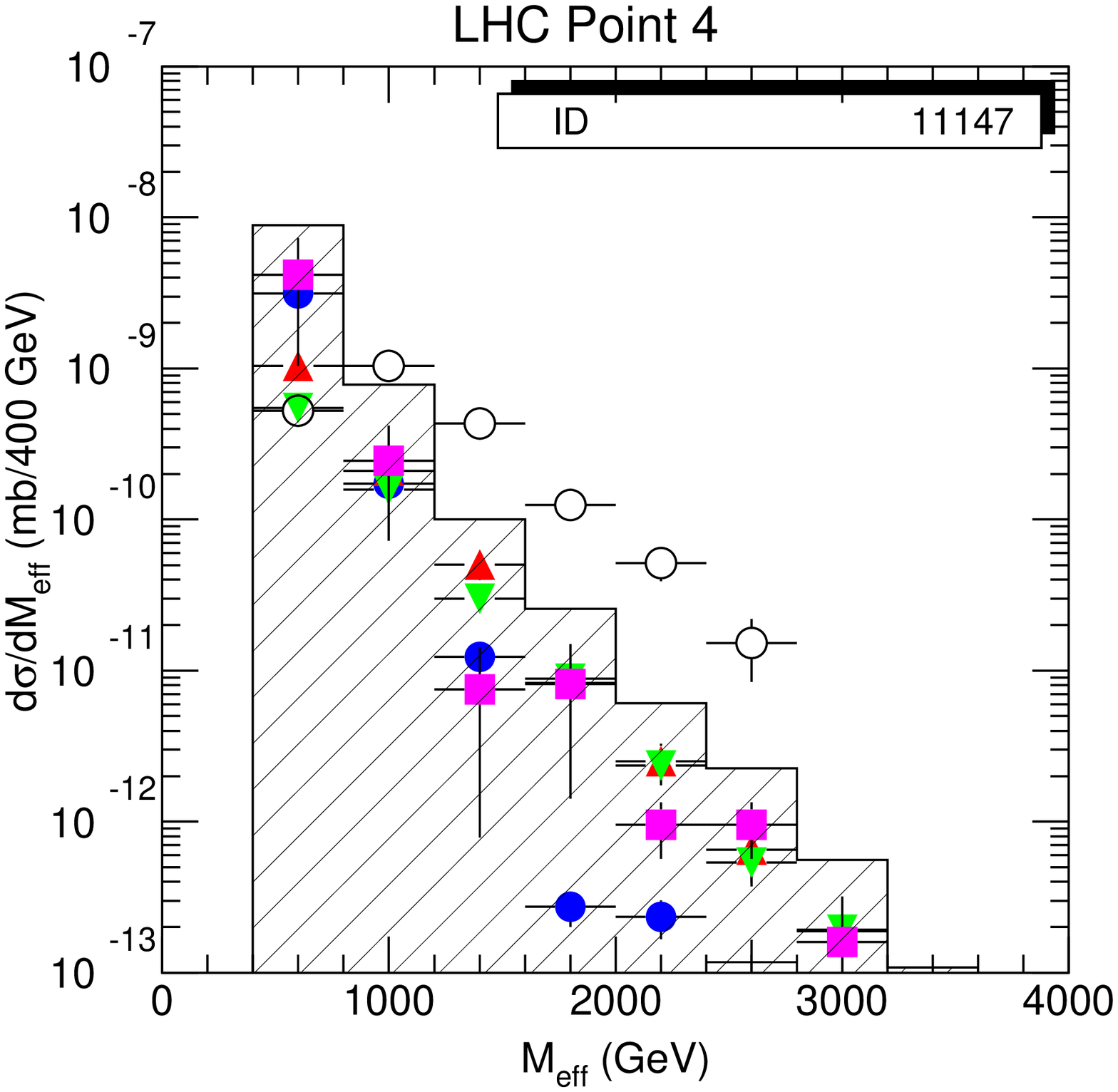}
\caption{SUSY signal and Standard Model backgrounds for LHC Point~4. See
Figure~1 for symbols.\label{point4_147}} 
\end{figure}

\begin{figure}[h]
\dofig{3.20in}{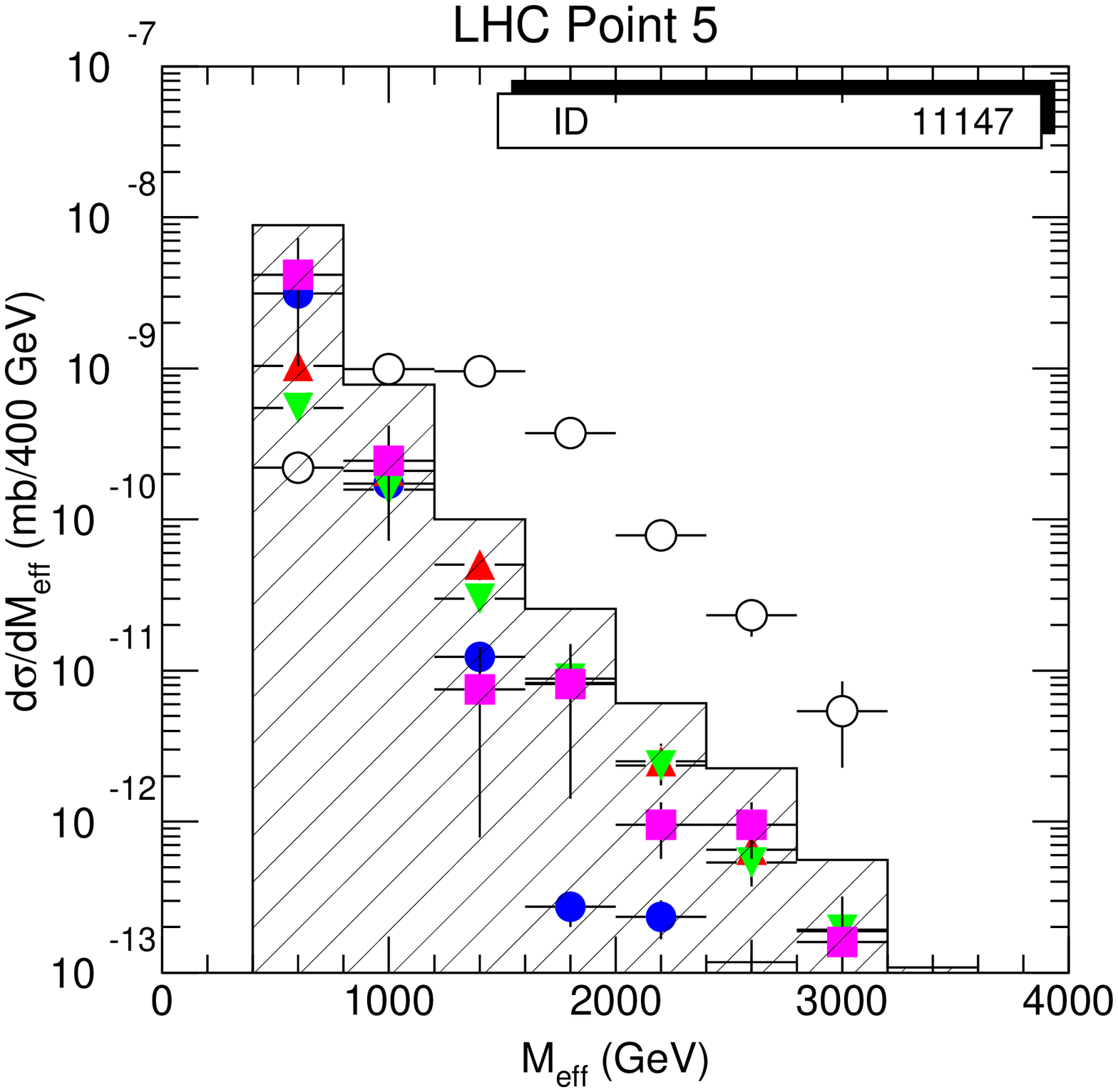}
\caption{SUSY signal and Standard Model backgrounds for LHC Point~5. See
Figure~1 for symbols.\label{point5_147}} 
\end{figure}

	The first step in the search for new physics is to discover a deviation from the Standard
Model and to estimate the mass scale associated with it. SUSY
production at the LHC is dominated by gluinos and squarks, which decay
into multiple jets plus missing energy. A variable which is sensitive
to inclusive gluino and squark decays is the effective mass $\Meff$,
defined as the scalar sum of the $p_T$'s of the four hardest jets and
the missing transverse energy $\etmiss$,
$$
\Meff = p_{T,1} + p_{T,2} + p_{T,3} + p_{T,4} + \etmiss\,.
$$
Here the jet $p_T$'s have been ordered such that $p_{T,1}$ is the
transverse momentum of the leading jet.
The Standard Model backgrounds tend to have smaller $\etmiss$, fewer
jets and a lower jet multiplicity.  In addition,
since a major source of $\etmiss$ is weak decays, large $\etmiss$ events in the Standard Model
tend to have the missing energy balanced by leptons.  To
suppress these backgrounds, the following cuts were made:  
\begin{itemize}
\item	$\etmiss > 100\,\GeV$;
\item	$\ge4$ jets with $p_T > 50\,\GeV$ and  $p_{T,1} > 100\,\GeV$;
\item	Transverse sphericity $S_T > 0.2$;
\item	No $\mu$ or isolated $e$ with $p_T > 20\,\GeV$ and $\eta<2.5$; 
\item	$\etmiss > 0.2 \Meff$.
\end{itemize}
 With these cuts and the idealized detector
assumed here, the signal for all five LHC points is much larger than
the Standard Model backgrounds for large $\Meff$, as is illustrated in
Figures.~\ref{point1_147}--\ref{point5_147}.

\begin{table}[b]
\caption{The value of $\Meff$ for which $S = B$ compared to $\Msusy$,
the lighter of the gluino and squark masses. Note that Point~3 is
strongly influenced by the $\etmiss$ and jet $p_T$ cuts.\label{meff}}
\begin{center}
\begin{tabular}{cccc}
\hline\hline
LHC Point& $\Meff\,(\GeV)$& $M_{\rm SUSY}\,(\GeV)$& Ratio\\
\hline
1 &      1360           &   926 &    1.47 \\
2 &      1420           &   928 &    1.53 \\
3 &      \phantom{0}470 &   300 &    1.58 \\
4 &      \phantom{0}980 &   586 &    1.67 \\
5 &      \phantom{0}980 &   663 &    1.48 \\
\hline\hline
\end{tabular}
\vskip-10pt
\end{center}
\end{table}

	The peak of the $\Meff$ mass distribution, or alternatively
the point at which the signal (S) begins to exceed the standard model background (B), provides a
good first estimate of the SUSY mass scale, which is defined to be
$$
\Msusy = \min(M_{\tg}, M_{\tilde u_R})
$$ 
While $\Msusy$ obviously should be set by the gluino and squark masses,
the choice of $M_{\tilde u_R}$ as the typical squark mass is somewhat
arbitrary. The ratio of the value $\Meff$ for which $S = B$ to
$\Msusy$ was calculated by fitting smooth curves to the signal and
background and is given in Table~\ref{meff}. It must be noted,
however, that for LHC Point 3 the cuts produce a minimum value of
$\Meff$ near the crossover. A more realistic treatment of the
$\etmiss$ resolution could be important for this point. At this point
event rates are so large that this step in our procedure is not needed;
we will not use $\etmiss$ in the analyses shown below.

\begin{figure}[h]
\dofig{3.20in}{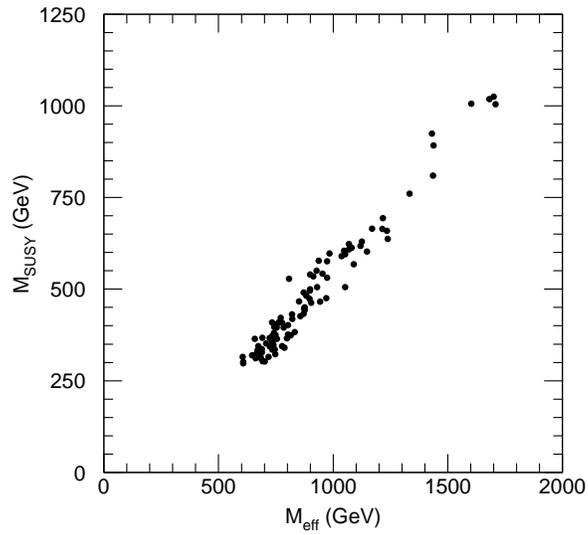}
\caption{Scatterplot of $\Msusy = \min(M_{\tg}, M_{\tilde u})$
vs.\ $\Meff$ for randomly chosen SUGRA models having the same light
Higgs mass within $\pm3\,\GeV$ as LHC Point~5.\label{scan1}}
\end{figure}

\begin{figure}[h]
\dofig{3.20in}{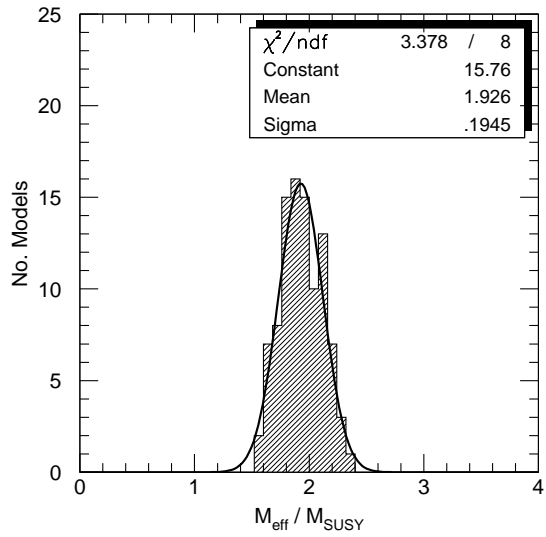}
\caption{Ratio $\Meff/\Msusy$ from Figure~\protect\ref{scan1}. The
distribution of this ratio is approximately Gaussian with a width
about 10\% of its mean.\label{scan3}} 
\end{figure}

	To see whether the approximate constancy of this ratio might
be an accident, 100 SUGRA models were chosen at random with $100 < m_0
< 500\,\GeV$, $100 < m_{1/2} < 500\,\GeV$, $-500 < A_0 < 500\,\GeV$,
$1.8 < \tan\beta < 12$, and $\sgn\mu=\pm1$.  These models
were  compared to the assumed
signal, LHC Point~5.  The light Higgs was assumed to be known, and all
the comparison models were required to have $M_h$ within $ \pm 3\,\GeV$
of its nominal value;
the 3~GeV error is determined not be the experimental measurements but
by an estimate of the theoretical uncertainties on the prediction of
$M_h$ from the parameters of the SUGRA model. 
A sample of 1K events
was generated for each point, and the peak of the $\Meff$ distribution
was found by fitting a Gaussian near the peak.  Figure~\ref{scan1}
shows the resulting scatterplot of $\Msusy$ vs.{} $\Meff$.  The ratio
is constant within about $\pm10\%$, as can be seen from
Figure~\ref{scan3}.  The mean value of the ratio is higher here than in
Table~\ref{meff} because this analysis uses the peak of the
distribution rather than the point at which it is equal to the
background. The error on the ratio is conservative, since there is
considerable contribution to the scatter from the limited statistics
and the rather crude manner in which the peak was found. While $\Meff$
does not provide a precise measurement, it has the advantage of being
generally applicable to a broad range of SUSY models. 

\newpage

\newpage

\newpage

\newpage

\newpage
\section{LHC Point 3: $m_0=200\,\GeV$, $\mhalf=100\,\GeV$, $\tan\beta=2$}
\label{sec:lhc3}

	LHC Point 3 has relatively light superpartners and hence a
very large production rate. At this point 200,000 events were
generated, corresponding to about 1 week of LHC running at low
luminosity; the statistical fluctuations on the plots are
due to this small Monte Carlo sample.  
All of the squarks of the first
two generations are heavier than the gluino, but one of the stop and
sbottom mass eigenstates is lighter than the gluino while the other is
heavier. The dominant SUSY production process is $\tilde{g}\tilde{g}$,
and the dominant gluino decay mode is $\tilde{g}\to \tilde{b} \bar b$.
The lightest $\tilde{b}$ is mainly $\tilde{b}_L$ and so decays
principally into $b\tchi_2^0$ since $\tchi_2^0$ ($\tchi_1^0$) is mainly
$\tilde{W}_3$ ($\tilde{B}$). Then $\tchi_2^0$ decays via virtual
sleptons to $\tchi_1^0 e^+e^-$ with a 16\% branching ratio. SUSY
events at this point are therefore dominated by final states involving
$b$-jets and pairs of opposite-sign, same-flavor leptons. Missing
transverse energy is not used in the analysis at this point.

\subsection{Measurement of $M_{\tchi_2}-M_{\tchi_1}$}

Events are selected by requiring:
\begin{itemize}
\item A pair of isolated leptons of opposite charge and the same flavor
with $p_{T\ell}>10$ GeV and $\abs{\eta_\ell}<2.5$;
\item At least two jets tagged as $b$ quarks and having $p_t>15$ GeV
and $\abs{\eta}<2$; a tagging efficiency of 60\% is assumed. 
\end{itemize}

\begin{figure}[h]
\dofig{3.20in}{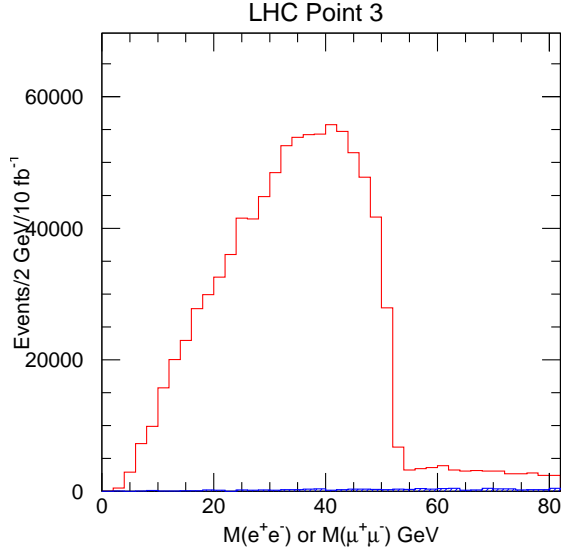}
\caption{The invariant mass distribution of $e^+e^-$ and $\mu^+\mu^-$
pairs arising at Point~3. The background, shown as a hatched histogram
is mainly due to $t\bar{t}$ events.\label{dilep-fig}} 
\end{figure}

The dilepton invariant mass distribution is shown in
Figure~\ref{dilep-fig}. The dominant Standard Model background is
$t\bar{t}$ production, which is quite small because it has smaller
color factors and requires two leptonic decays. This background, as
well as the combinatorial background from events with two $\tchi$
decays, can be eliminated by calculating the subtracted distribution:
$$
\left.{d\sigma\over dM}\right\vert_{\rm sub} = 
\left.{d\sigma\over dM}\right\vert_{e^+e^-}
+\left.{d\sigma\over dM}\right\vert_{\mu^+\mu^-}
-\left.{d\sigma\over dM}\right\vert_{e^+\mu^-}
-\left.{d\sigma\over dM}\right\vert_{e^-\mu^+}\,.  
$$
This subtracted mass distribution has a sharp edge at
$M_{\ell^+\ell^-}=M_{\tchi_2}-M_{\tchi_1}$, enabling this mass
difference to be measured with great precision. In view of the
enormous size of the event sample, the uncertainty on this measurement will
be limited by systematic effects. The large sample of $Z\to
\ell^+\ell^-$ decays will be used for calibration both of the mass
scale and of the relative $e$ and $\mu$ acceptance. The methods
employed will be similar to those used by CDF and D0 in their
determinations of the $W$ mass \cite{cdf-W,d0-W}. An 
estimate of 50~MeV for the uncertainty  on
$M_{\tchi_2}-M_{\tchi_1}$ should be conservative.

\subsection{Gluino and Sbottom Reconstruction}

The next step is a reconstruction of the gluino and sbottom masses by
combining a dilepton pair near the mass edge with jets. Events are
selected that have 
\begin{itemize}
\item At least two jets, tagged as having a $b$ quark with $p_t>15$
GeV and $\abs{\eta}<2$; a tagging efficiency of 60\% is assumed;
\item A $e^+e^-$ pair with $45$ GeV $<M_{\ell^+\ell^-}<55$ GeV and no
other electrons or a $\mu^+\mu^-$ pair in the same mass range and no
other muons in the event.
\end{itemize}
Since the mass of the  lepton pair is near its maximum value, 
in the rest frame of
$\tchi_2$ both $\tchi_1$ and the $\ell^+\ell^-$ pair are forced to be at
rest. The momentum of $\tchi_2$ in the laboratory frame is then
determined to be
$$
\vec P_{\tchi_2}=\left(1+M_{\lsp}/M_{\ell^+\ell^-}\right) 
\vec P_{\ell^+\ell^-}\,.
$$ 
where $M_{\lsp}$ must be assumed (see below).
This momentum can be combined with a $b$-jet to determine
$m_{\tilde{b}}$ and a second $b$-jet to determine $m_{\tilde{g}}$.
The $b$-jet energy and momentum must be corrected for the fact
that particles are lost outside the $R=0.4$ jet cone and for the
fact that weak decays produce neutrinos in the jets.  In this study, the correction
factor was determined using the data generated for LHC Point 5,
where the Higgs peak  ($h \ra b \bar b$) is observable.
In practice, techniques similar to those of references~\cite{cdf-top,d0-top}
would be used at LHC.

\begin{figure}[h]
\dofig{3.20in}{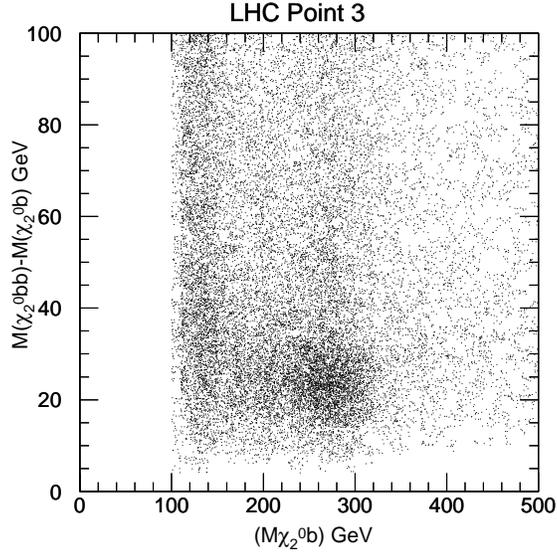}
\caption{The reconstruction of gluino and sbottom decays from the
decay chain $\tilde{g}\to \tilde\chi_2 (\to\tilde\chi_1 \ell^+\ell^-)
\tilde{b}$. Events are selected near the endpoint of the
$\ell^-\ell^+$ mass distribution (mass between 45 and 55 GeV) and the
momentum of $\tchi_2$ reconstructed. Two $b$-jets are then required and
the mass of $b+\tchi_2$ ($m=m_{\tilde{b}}$) and the mass difference
$\delta m=m_{bb\tchi_2}-m_{b\tchi_2}$ is computed. The scatterplot in
these two variables is shown. The $b$-jet energies have been
recalibrated and a tagging efficiency of 60\% per $b$ included.
\label{point4-gluino}}
\end{figure}

\begin{figure}[h]
\dofig{3.20in}{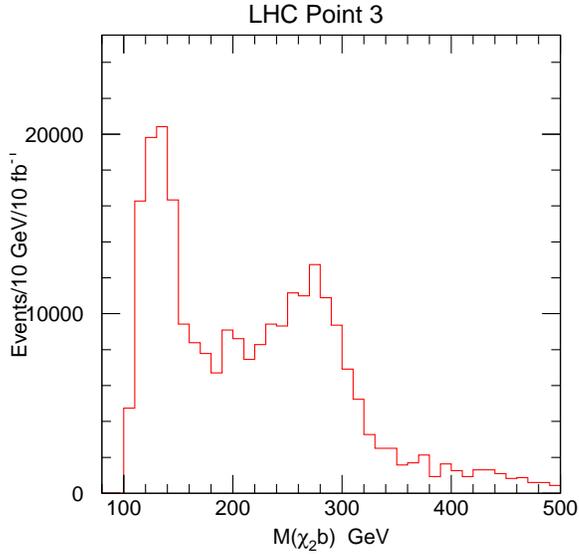}
\caption{The $M(\tilde b)$ projection of
Figure~\protect\ref{point4-gluino}.\label{x4sbottom} }
\end{figure}

\begin{figure}[h]
\dofig{3.20in}{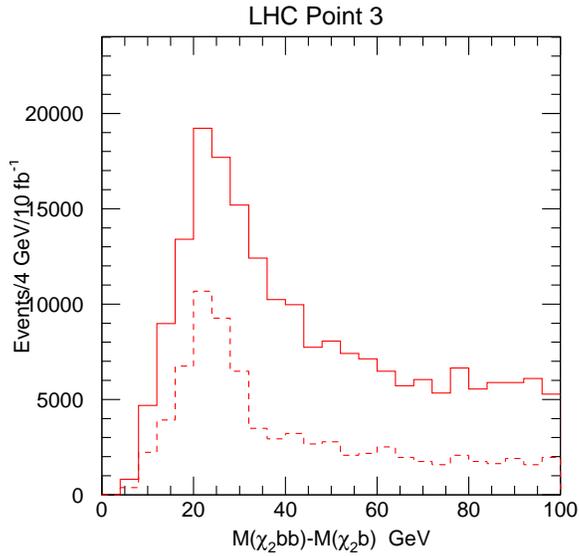}
\caption{The $M(\tg)-M(\tilde b)$ projection of
Figure~\protect\ref{point4-gluino}.\label{x4diff}
The dashed histogram shows the projection if a cut is made requiring that
the events lie in a slice of on the abscissa of  between 230 and 330 GeV of
Figure~\protect\ref{point4-gluino}.}
\end{figure}

Figure~\ref{point4-gluino} shows a scatterplot of
$m_{\tilde{g}}-m_{\tilde{b}}$ {\it vs.{}} $m_{\tilde{g}}$. Projections
onto the axes, shown in Figures~\ref{x4sbottom} and \ref{x4diff}, have
clear peaks. The positions of the peaks determine
$m_{\tilde{g}}-m_{\tilde{b}}$ and $m_{\tilde{b}}$ assuming that
$M_{\lsp}$ is known. Again, statistical errors are small and
the dominant errors will be from the determination of the jet energy
scale. A careful jet energy calibration has not been performed, so the
peaks in Figures.~\ref{x4sbottom} and \ref{x4diff} are displaced slightly
from their nominal values of 277.8 and 20.3 GeV. These systematic
errors can be estimated from those currently obtained by CDF and D0 in
the determination of the top quark mass \cite{cdf-top,d0-top}.  
The mass difference $m_{\tilde{g}}-m_{\tilde{b}}$ is insensitive to
the assumed $\lsp$ mass while the reconstructed sbottom peak moves. 

\begin{figure}[h]
\dofig{3.20in}{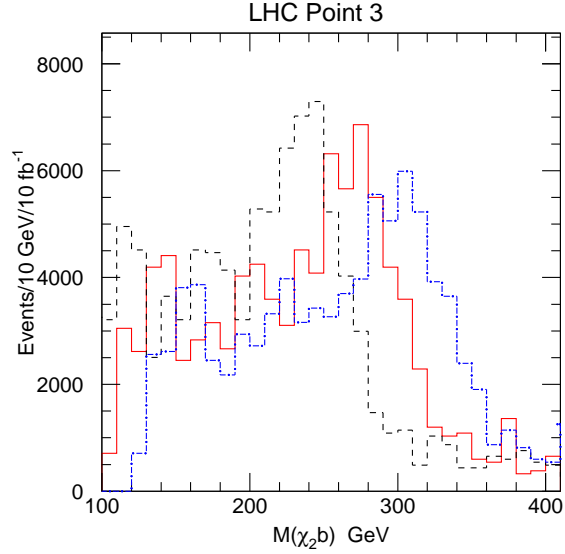}
\caption{The same as Figure~\protect\ref{x4sbottom} with the addition
of two more histograms (dashed and dotted) showing the result if the
assumed value of $m_{\tchi_i^0}$ is varied by $\pm 20 $ GeV. A cut is
imposed on the mass difference $\abs{m_{\tilde{g}}-m_{\tilde{b}}-20}$
GeV $<15$ GeV before the projection of the scatterplots is made. 
\label{gluino-moving}}
\end{figure}

The dependence of the $\tilde b$ mass peak on the assumed value of
$M_{\lsp}$ is shown in Figure~\ref{gluino-moving}, where $M_{\lsp}$ is
varied by $\pm 20\,\GeV$ from its nominal value.  In making this plot
we have required that the mass difference
$M(\tchi_2^0 bb)-M(\tchi_2^0 b)$ be within 15 GeV of the value where its distribution peaks.
  This cut
removes considerable background as can be seen by comparing the peaks
in this figure with that in Fig\ref{x4sbottom}. We estimate
$$
M_{\tilde{b}}({\rm measured})-M_{\tilde{b}}({\rm true})=
1.5\Bigl(M_{\lsp}({\rm assumed})-M_{\lsp}({\rm true})\Bigr) \pm 3\,\GeV
$$
and
$$
M_{\tilde{g}}({\rm measured})-M_{\tilde{b}}({\rm measured})=
M_{\tilde{g}}({\rm true})-M_{\tilde{b}}({\rm true})\pm 2\,\GeV
$$ 
The $\lsp$ mass will be determined by a global fit of the SUSY model
to all the measurements; see Section~\ref{sec:scan}.

\subsection{Light Squark Reconstruction}

Light squarks can also be reconstructed at this point using the decay
chain $\tilde{q_L}\to \tchi_2^0 q$, which has a branching ratio of
approximately 10\%. There is an enormous background from gluino decays
to $\tilde{b}b$, so events must be rejected if there is a $b$-jet
present.  We use the ATLAS $b$-tagging study (see Figure~3.42 of
Ref.\cite{ATLAS}).  At low luminosity this study implies that a
tagging efficiency of 90\% for $b$-jets can be achieved at the price
of misidentifying 25\% of the light quark jets as $b$-jets.  While
this mistag rate is not adequate in the cases where a $b$-tag is required,
it implies that 90\% of the $b$-quark jets can be vetoed and 75\% of
the light quark jets accepted by the same cut. This veto prescription
is used in this subsection.

\begin{figure}[h]
\dofig{3.20in}{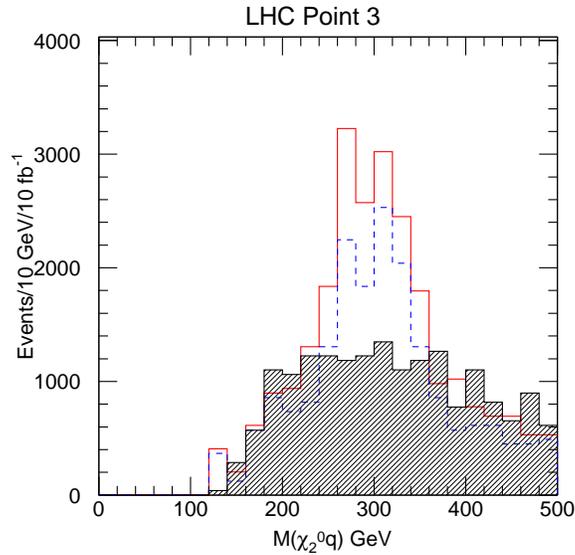}
\caption{Reconstructed $\tilde{q_L}$ mass at Point~3. The combinatorial
background estimate is shown as a hatched histogram and the events due
to light squarks as the dashed histogram. The remaining events are due
to gluino decays where a $b$-jet is misidentified as a light quark jet.
\label{squarkpointc3}}
\end{figure}

Events are selected as follows:
\begin{itemize}
\item At least one jet with $p_t>125$ GeV and $\abs{\eta}<2$.
\item No $b$-jets with $p_t>15$ GeV and $\abs{\eta}<2$; a vetoing
efficiency of 90\% is assumed and 25\% of non $b$-jets are assumed to
be rejected also.
\item An $e^+e^-$ pair with $45$ GeV $<M_{\ell^+\ell^-}<55$ GeV and no
other electrons or a $\mu^+\mu^-$ pair in the same mass range and no
other muons in the event.
\end{itemize}
The reconstruction of the momentum of $\tilde\chi_2^0$ is performed
using the same method as above by selecting events near the endpoint of
the dilepton mass distribution.  We assume that the SUGRA model is
used to infer the mass of $\tilde\chi_1^0$ from the
$\tilde\chi_2^0-\tilde\chi_1^0$ mass difference. Jets of
$\abs{\eta}<2$ and $p_t>125\,\GeV$ are now combined with the
$\tilde\chi_2^0$ and the mass distribution is shown in
Figure~\ref{squarkpointc3}.  Even with the 90\% vetoing efficiency for
$b-$quarks there are a significant number of $b-$jets remaining in
this plot.  The contribution from the light squarks is shown as the
dashed-histogram.  If the vetoing efficiency were raised to 95\%
approximately one-half of the remaining $b-$jets are removed and
consequently the peak moves to a larger mass.  The peak shown has
contributions from $\tilde{b_L}$ of mass  $278 $ GeV and the light
quarks that have mass around 310 GeV.  Charge $-1/3$ and $+2/3$ squarks are separated by about 5 GeV
in mass; this contributes to the broadening of the peak. That the peak
is real can be seen be estimating the combinatorial background as
follows. Events are mixed by taking the $\tilde\chi_2^0$ momentum from
one event and the jet from another; both events satisfying the same
selection criteria. The mass distribution obtained in this way is
shown as the hatched distribution in Figure~\ref{squarkpointc3}.
Conservatively, we estimate an error of 20 GeV on the average
$\tilde{q}_L$ mass from this method.

\subsection{Branching ratio of $\tchi_2\to \tchi_1^0\ell^+\ell^-$}
\label{sub:br}
By selecting events with four tagged $b-$jets and either two or four 
isolated leptons,
the product of branching 
ratios $BR(\tilde{\chi_2^0} \to \lsp \ell^+\ell^-)\times
 BR(\tilde{b}\to b \tilde{\chi_2^0} X)$ can be determined. 
There are
150000 events/10~fb$^{-1}$ with two dilepton pairs and four $b-$jets. The backgrounds from non supersymmetric sources 
are
negligible, and again therefore the dominant uncertainties are systematic. 
Using a value of $3\% $ for the uncertainty on the absolute
lepton acceptance, we expect that 
$BR(\tilde{\chi_2^0}\to \lsp e^+e^-)\times BR(\tilde{b}\to b \tilde{
\chi_2^0} X)$ can be determined to be $(14.0 \pm 0.5)\%$

\subsection{Electroweak Production of Superpartners}

At this SUGRA point, sleptons cannot be produced from the decay of strongly
interacting sparticles. The production rates are therefore quite small
despite the low masses ($m_{\tilde{e}_L}=215$ GeV,
$m_{\tilde{e}_R}=206$ GeV) as they must be pair produced in Drell-Yan
like processes. The heavier charginos and neutralinos are only rarely
produced in the decays of gluinos, so again their dominant production
mechanism is electroweak. Unlike the case of sleptons,
the direct production rate of the lighter
charginos and neutralinos is quite large.  An attempt has been made to
isolate these processes.  This is an example of a case where the
analysis of a complete SUSY signal is needed.  The signals that we are
attempting to extract stand clearly above Standard Model backgrounds,
but we face the large background from the production of strongly
interacting sparticles.  As so few events pass the cuts, we generated
separate data samples corresponding to the electroweak production of
sparticles and reweighted the events appropriately.

\begin{figure}[h]
\dofig{3.20in}{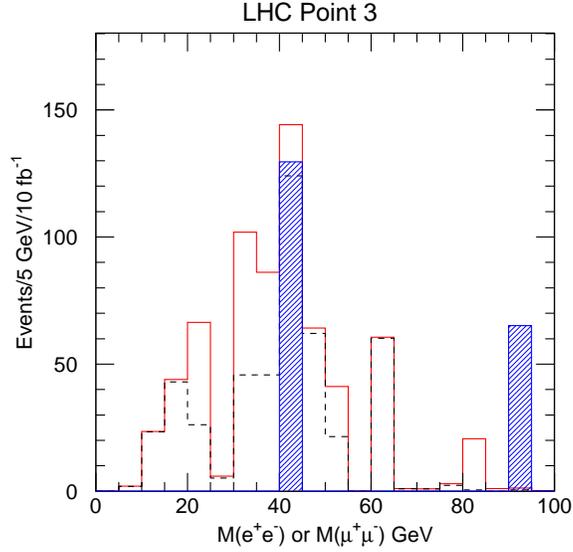}
\caption{The invariant mass distribution of $e^+e^-$ and $\mu^+\mu^-$
pairs arising at Point~4. Events are selected requiring no jets with
$p_t>30\,\GeV$ in $\abs{\eta}<3$ and at least three isolated leptons,
two of which are of the same flavor and opposite charge. Lepton
detection efficiency of 90\% per lepton is included.  The dashed
histogram shows the contribution arising from the direct production of
$\tchi_1^+\tchi_2^0$ final states. The background is shown as the hatched
histogram.\label{winoend}}
\end{figure}

Events are selected that have:
\begin{itemize}
\item A three isolated leptons a pair of which have opposite charge
and the same flavor with $p_{T\ell}>10$ GeV and $\abs{\eta}<2.5$;
\item No jets with $p_t> 30\,\GeV$ in $\abs{\eta}<3.0$.
\end{itemize}
The jet veto is needed to remove gluino and squark initiated events.
These events have jets in the central region arising from the decay
products of the sparticles and from final state gluon radiation. These
events also have jets, approximately uniform in rapidity, from initial
state radiation.  This latter source is also present in the direct
production of charginos, neutralinos and sleptons.
Figure~\ref{winoend} shows the dilepton invariant mass distribution of
the two leptons that have opposite charge and the same flavor.  The
number of {\it generated} events in this plot is not large, but are
sufficient to demonstrate that in $10$~fb$^{-1}$ of data there will be
sufficient events for a precise measurement.  The background events in
this plot (corresponding to three generated events) are from
$t\overline{t}$ production, the third lepton being from the decay of a
$b$-quark. A stricter jet veto (20~GeV instead of 30~GeV) reduces this
background further.  

There is an indication of an edge in the mass distribution
corresponding to the decay $\tchi_2^0\to\tchi_1^0\ell^+\ell^-$. The
events in this plot are dominated by the production of
$\tchi_2^0\tchi_1^\pm$ final states whose contribution is shown as the
dotted histogram.  If two isolated leptons are required and the same
plot made the result is that there are more events. There is now a potential
background from Drell-Yan production of dilepton events which must be
eliminated by a cut on missing $\etmiss$ or the angle between the two
leptons; the Drell-Yan events are back-to-back while in the SUSY
events the leptons arise from $\tchi_2^0\to\tchi_1^0\ell^+\ell^-$ and
are therefore close in angle.  The production rates in these two and
three lepton final states can be compared and used to provide a powerful argument
concerning the origin of the lepton samples and provide an additional
constraint on the model since, as we will demonstrate in
section~\ref{sec:scan}, the measurements that have been made using the
strong production of sparticles fix the model parameters, resulting in
a {\it prediction} for the rates shown in Figure~\ref{winoend}.

In principle, the decay $\tilde{e}_L\to\tchi_2^0 e$ should be
reconstructible by selecting with a least 3 isolated leptons, an
oppositely charged pair of which have mass between 45 and 55 GeV.  The
momentum of $\tchi_2^0$ is reconstructed as above and then combined
with a third lepton to search for a reconstructed $\tilde{e}_L$.  The
extraction of this signal is very difficult. The production rate for
gauginos provides a serious background that can only be controlled by
increasing the number of isolated leptons required. The dominant
slepton production process is $\tilde{e_L}+\tilde{\nu_e}$. This can be
extracted only by requiring at least {\it four} isolated leptons from
the decay chain
$$
\begin{array}{lll}
\tilde\ell_L^+ &+& \tilde\nu_\ell \\
\downarrow && \downarrow \\
\tchi_2^0+\ell^+ && \tchi_1^+ +\ell^- \\
\downarrow && \downarrow \\
\lsp +\ell^+ +\ell^- && \lsp +\ell^+ +\nu
\end{array}
$$
or alternatively from 
$$
\begin{array}{lll}
\tilde\ell_L^+ &+& \tilde\nu_\ell \\
\downarrow && \downarrow \\
\tchi_2^0+\ell^+ && \tchi_2^0+\nu \\
\downarrow && \downarrow \\
\lsp +\ell^+ +\ell^- && \lsp +\ell^+ +\ell^-
\end{array}
$$
The dominant decay chain $\tilde\nu_\ell\to\tchi_1^+\ell$, $\tchi_1^+\to
\lsp+{\rm jets}$ is killed by the jet veto requirement. The experiment
is only feasible at high luminosity.

\newpage

\newpage

\newpage

\newpage

\newpage
\section{LHC Point~4: $m_0=800\,\GeV$, $\mhalf=200\,\GeV$, $\tan\beta=10$}
\label{sec:lhc4}

	LHC Point~4 has squarks which are much heavier than gluinos, so
production of the latter is dominant. The heavier chargino and
neutralinos have a much larger admixture of gauginos than at the other
points. Hence, the gluinos decay into all combinations of all the
charginos and neutralinos and all the quark pairs with comparable
branching ratios, giving a very complex mixture of signatures.

\subsection{Selection of gaugino decays $\tchi_i \ra \tchi_j \ell^+\ell^-$}

\begin{figure}[h]
\dofig{3.20in}{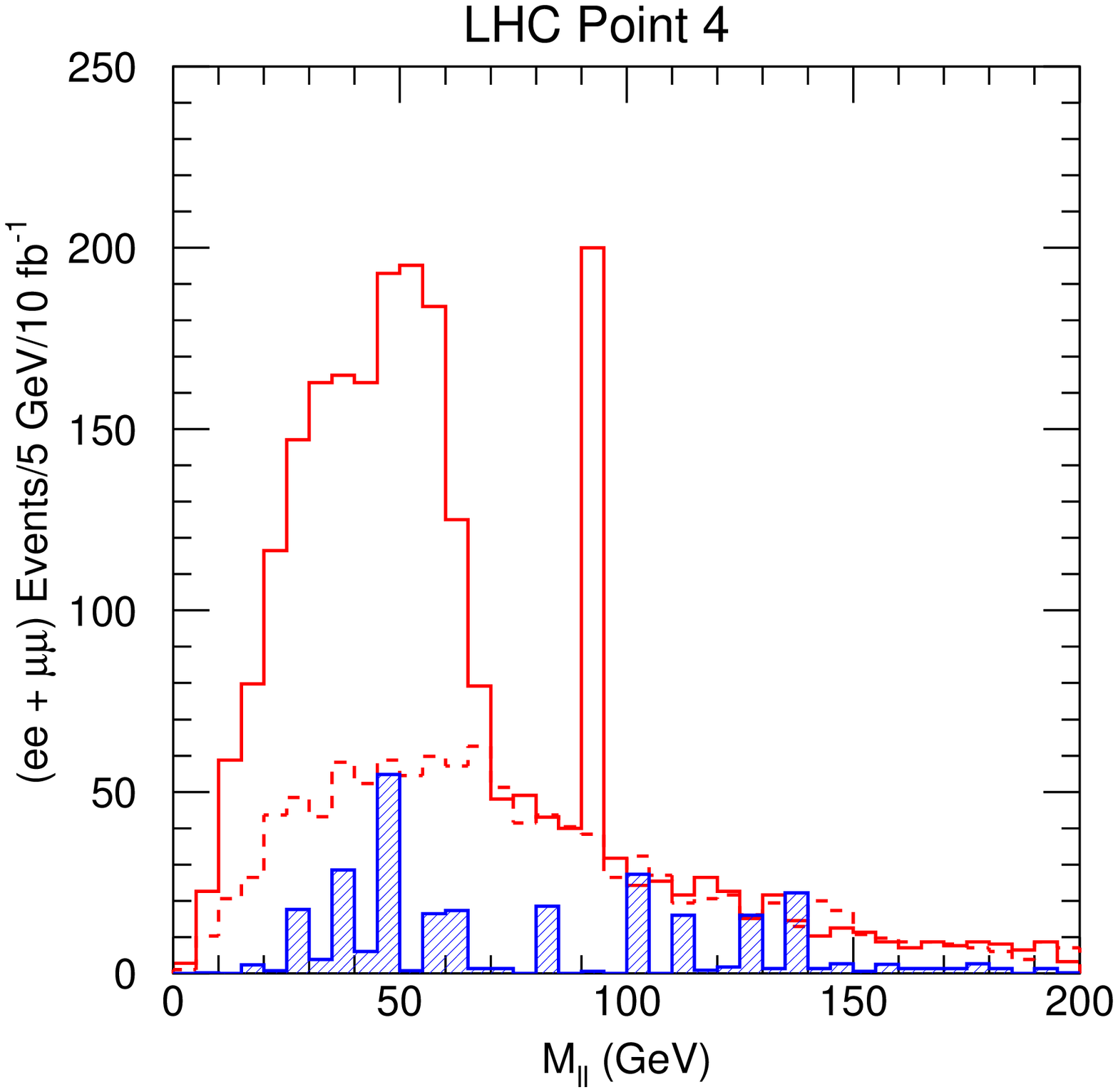}
\caption{$M_{\ell^+\ell^-}$ distribution for opposite-sign,
same-flavor dileptons for the Point~4 signal (solid histogram),
opposite-sign, opposite-flavor dileptons (dashed histogram), and
Standard Model background (shaded histogram).\label{c4_mllos}}
\end{figure}

	The objective of this analysis is to isolate opposite-sign,
same-flavor dileptons coming from $\tchi_2^0 \ra \lsp \ell^+\ell^-$,
$\tchi_{3,4}^0 \ra \lsp Z$, and $\tchi_2^\pm \ra \tchi_1^\pm Z$.
First, the following cuts were made to suppress the Standard Model
backgrounds: 
\begin{itemize}
\item	$\Meff > 800\,\GeV$;
\item	$\etmiss > \max(100\,\GeV,0.15\Meff)$;
\item	$\ge 4$ jets  with $p_{T,1} > 100\,\GeV$, $p_{T,2,3,4} >
50\,\GeV$;
\item	$\ell^+\ell^-$ pair with $p_{T,\ell}> 10\,\GeV$, $\eta_\ell <
2.5$;
\item	$\ell$ isolation cut: $E_T < 10\,\GeV$ in $R=0.2$;
\item	Transverse sphericity $S_T > 0.2$.
\end{itemize}
The opposite-sign, same-flavor and opposite-flavor dilepton mass
spectra for the Point~4 signal and the opposite-sign, same-flavor
Standard Model background with these cuts are shown in
Figure~\ref{c4_mllos}. There are clear low-mass and $Z$ opposite-sign,
same-flavor signals. (Note that the $Z$ is treated as a narrow
resonance in the event generator.)  The same-sign Standard Model
background is not shown but is smaller than the opposite-sign
background.

\begin{figure}[h]
\dofig{3.20in}{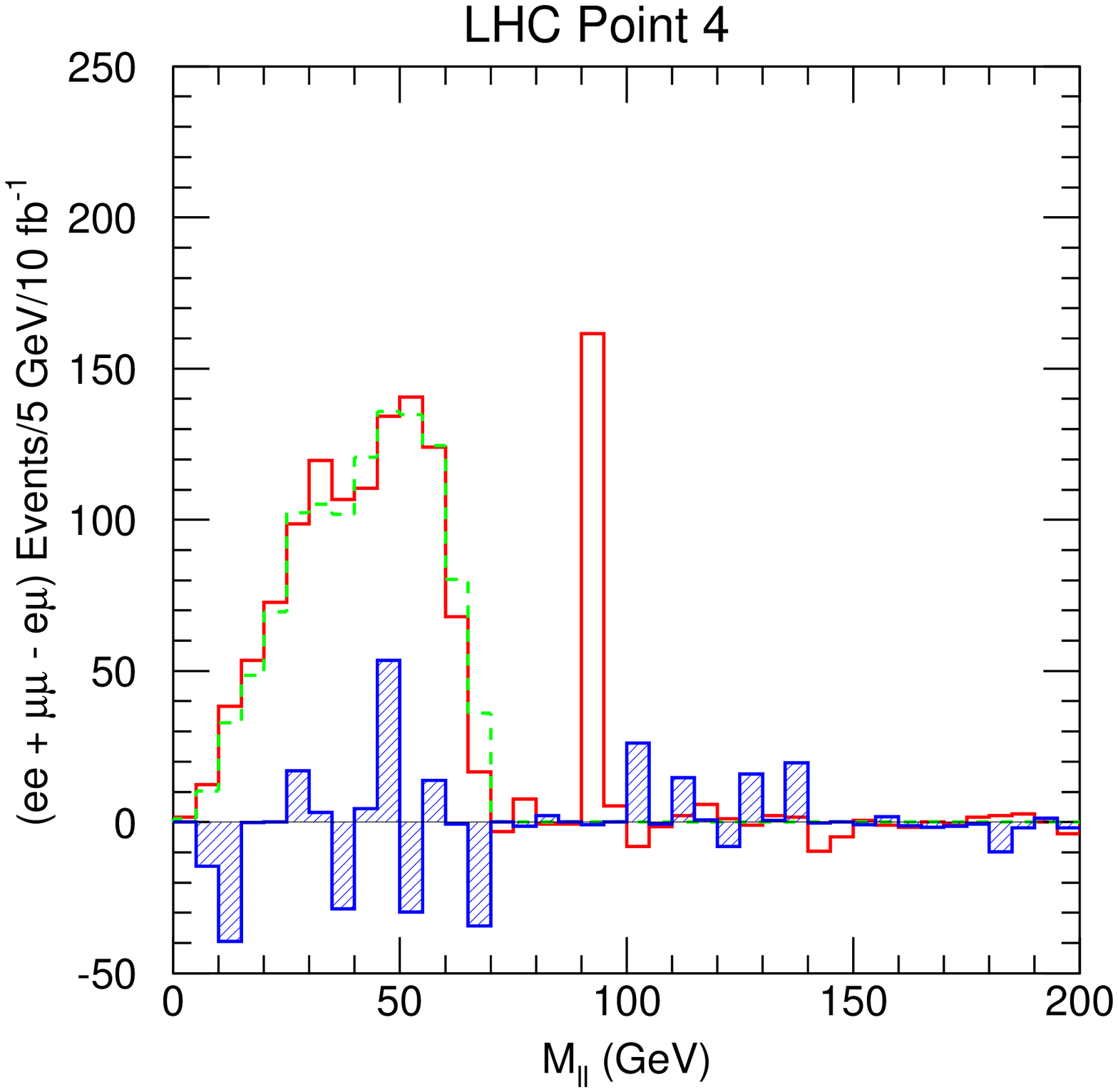}
\caption{Difference of the $M_{\ell^+\ell^-}$ distribution for
opposite-sign, same-flavor dileptons and opposite-sign, opposite-flavor
dileptons for the Point~4 signal (open histogram) and the Standard Model
background (shaded histogram).\label{c4_mllosdif}}
\end{figure}

	The difference of the opposite-sign, same-flavor and
opposite-sign, opposite-flavor dilepton distributions is shown in
Figure~\ref{c4_mllosdif}. This difference should only have
contributions from $\tchi_2^0 \ra \lsp \ell^+\ell^-$ (shown as a
dashed curve) and from $Z \ra \ell^+\ell^-$ decays from heavy
charginos and neutralinos, which contribute to the $Z$ peak.
Contributions from two independent chargino, top, or $W$ decays contribute
equally to both flavor combinations and therefore cancel in this figure. The
Standard Model background in the figure fluctuates in sign because of
limited statistics but should also mostly cancel.

	The edge of the $\tchi_2^0 \ra \lsp \ell^+\ell^-$ signal is
not quite as sharp as in previous cases, but it clearly can be
measured with an error of $\sim1\,\GeV$ or less. The observation both
of this edge and of the $Z$ peak shows that both light and heavy
gauginos contribute, since for any given gaugino $\tchi_i$, decay into
$\tchi_j Z$ is much larger than decay into $\tchi_j \ell^+\ell^-$. 

	The relative number of events in the two parts of the
distribution can be measured with a statistical error of a few
percent. The systematic error on the $e$ and $\mu$ acceptance should
be comparable to or 
less than this after $Z$ decays are studied carefully and used for 
calibration of the calorimeter. Since the
sleptons are also heavy at this point, the leptonic branching ratios
for the $\tchi_i^0$ are essentially determined by the $Z$ branching
ratios, so the relative number of events provides a measure of
$$
\sum\limits_{\tilde\chi_i = \tilde\chi_{3,4}^0,\tilde\chi_2^\pm}
B(\tilde g \ra \tilde\chi_i X) B(\tilde\chi_i \ra Z X) \over
B(\tilde g \ra \tilde\chi_2 X) 
$$
There are of course non-negligible corrections from squark production
and from lepton acceptance.

\begin{figure}[h]
\dofig{3.20in}{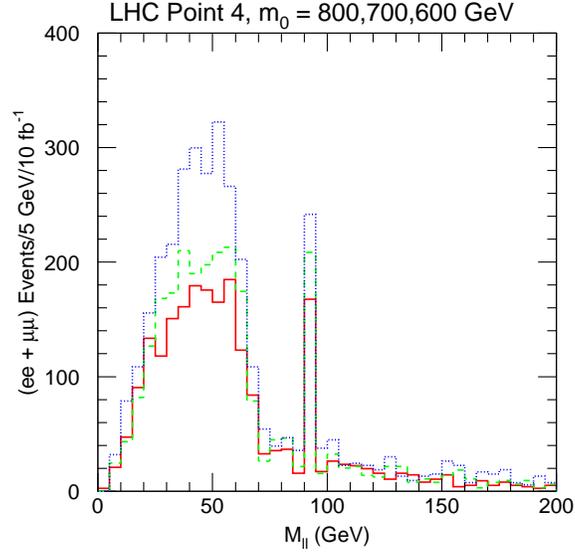}
\caption{Comparison of $M_{\ell^+\ell^-}$ distributions for
opposite-sign, same-flavor dileptons for $m_0=800\,\GeV$ (solid
histogram), $m_0=700\,\GeV$ (dashed histogram), and $m_0=600\,\GeV$
(dotted histogram).\label{c4_mllm0}}
\end{figure}

\begin{figure}[h]
\dofig{3.20in}{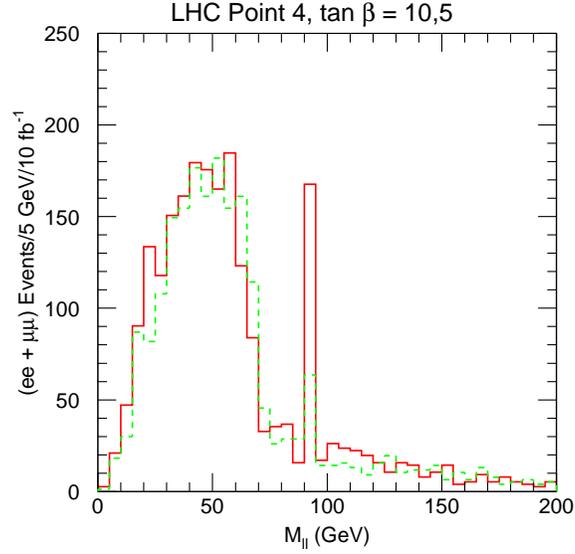}
\caption{Comparison of $M_{\ell^+\ell^-}$ distributions for
opposite-sign, same-flavor dileptons for $\tan\beta=10$ (solid
histogram) and $\tan\beta=5$ (dashed histogram).\label{c4_mllbeta}}
\end{figure}

	To see how useful such a branching ratio measurement might be,
samples of 10K events each with $m_0 = 800$, 700, and $600\,\GeV$ were
generated, forcing the decays $\tchi_2^0 \ra \lsp \ell^+\ell^-$ and $Z
\ra \ell^+\ell^-$. Figure~\ref{c4_mllm0} shows the resulting
opposite-sign, same-flavor mass distributions including the 6\%
leptonic branching ratios. (Subtraction of the same-sign background
does not work properly when decays are forced, so the subtracted
distribution is not shown.) The number of events in the $Z$ peak and
below 70~GeV are 210/2000, 260/2350, and 3000/3150 respectively. The
ratio is nearly constant, and the change in absolute number is similar
to the change in the total cross section. There is, however,
sensitivity to $\tan\beta$, which is more directly related to the
mixing of the heavy and light gauginos. A sample of 10K events with
$m_0=800\,\GeV$ and $\tan\beta=5$ was generated, again forcing the
decays. Figure~\ref{c4_mllbeta} shows the two distributions; the ratio
is 80/1950. More study is needed, but it seems likely that this ratio
could constrain $\tan\beta=10$ to $\sim10\%$.

\begin{figure}[h]
\dofig{3.20in}{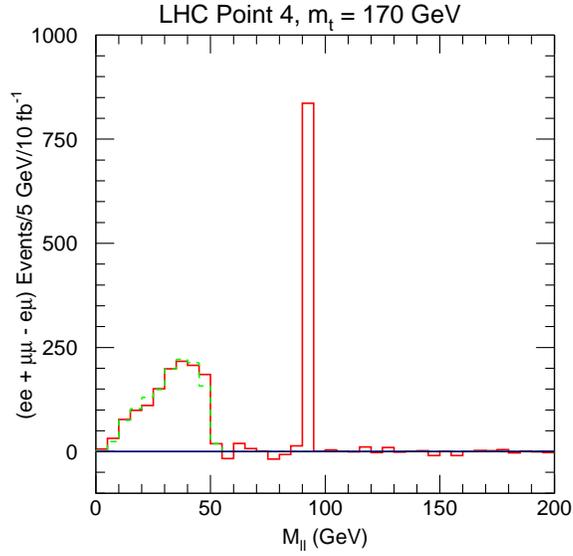}
\caption{The same as Figure~\protect\ref{c4_mllosdif} but with
$m_t=170\,\GeV$. Note the large change in the size of the $Z$ peak.
\label{c4_mllosdif170}}
\end{figure}

	The results at this point are very sensitive to the top quark
mass. For example, Figure~\ref{c4_mllosdif170} shows the same
distribution as Figure~\ref{c4_mllosdif} for the same SUSY parameters
but for $m_t = 170\,\GeV$ instead of $175\,\GeV$. Note that the $Z$
peak is dramatically larger. The reason for this extreme sensitivity
is that this point is very close to the boundary of the allowed
region: there is no electroweak symmetry breaking for $m_t=165\,\GeV$.
The possibility of such sensitivity, however, suggests that 
theoretical uncertainties could play an important role in this region
of parameter space.

\subsection{Selection of $\tilde\chi_4^0 \ra \tilde\chi_1^\pm W^\mp
\ra e^\pm \mu^\mp X$}

	Isolated $e\mu$ events come from two independent $W$ or wino
decays, not from single $Z$ or neutralino decays. At LHC Point~4 the
branching ratio for $\tilde\chi_4^0 \ra \tilde\chi_1^\pm W^\mp$ is
about 84\%; this decay contributes to $e^\pm \mu^\mp$ but not to
$e^\pm \mu^\pm$. Since the gluino is a Majorana fermion, other
channels involving two independent $\tg\tg$ or $\tg\tq$ decays
contribute equally to $e^\pm \mu^\mp$ and $e^\pm\mu^\pm$.

\begin{figure}[h]
\dofig{3.20in}{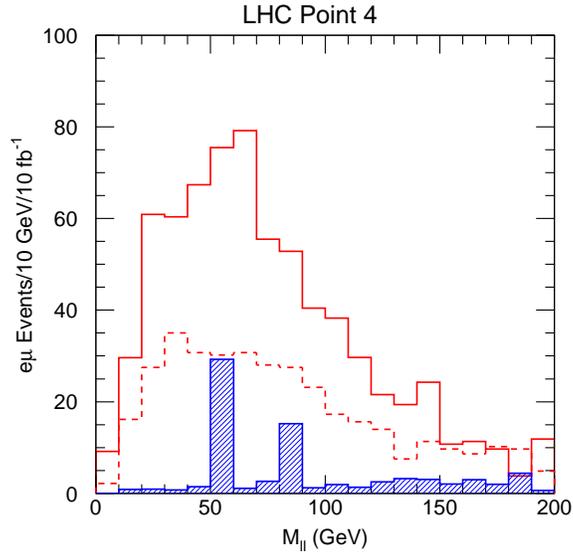}
\caption{$M_{e^\pm\mu^\mp}$ (solid) and $M_{e^\pm\mu^\pm}$ (dotted)
distributions for the Point~4 signal, and Standard Model opposite-sign,
opposite-flavor background (shaded histogram).\label{c4_mllemu}}
\end{figure}

\begin{figure}[h]
\dofig{3.20in}{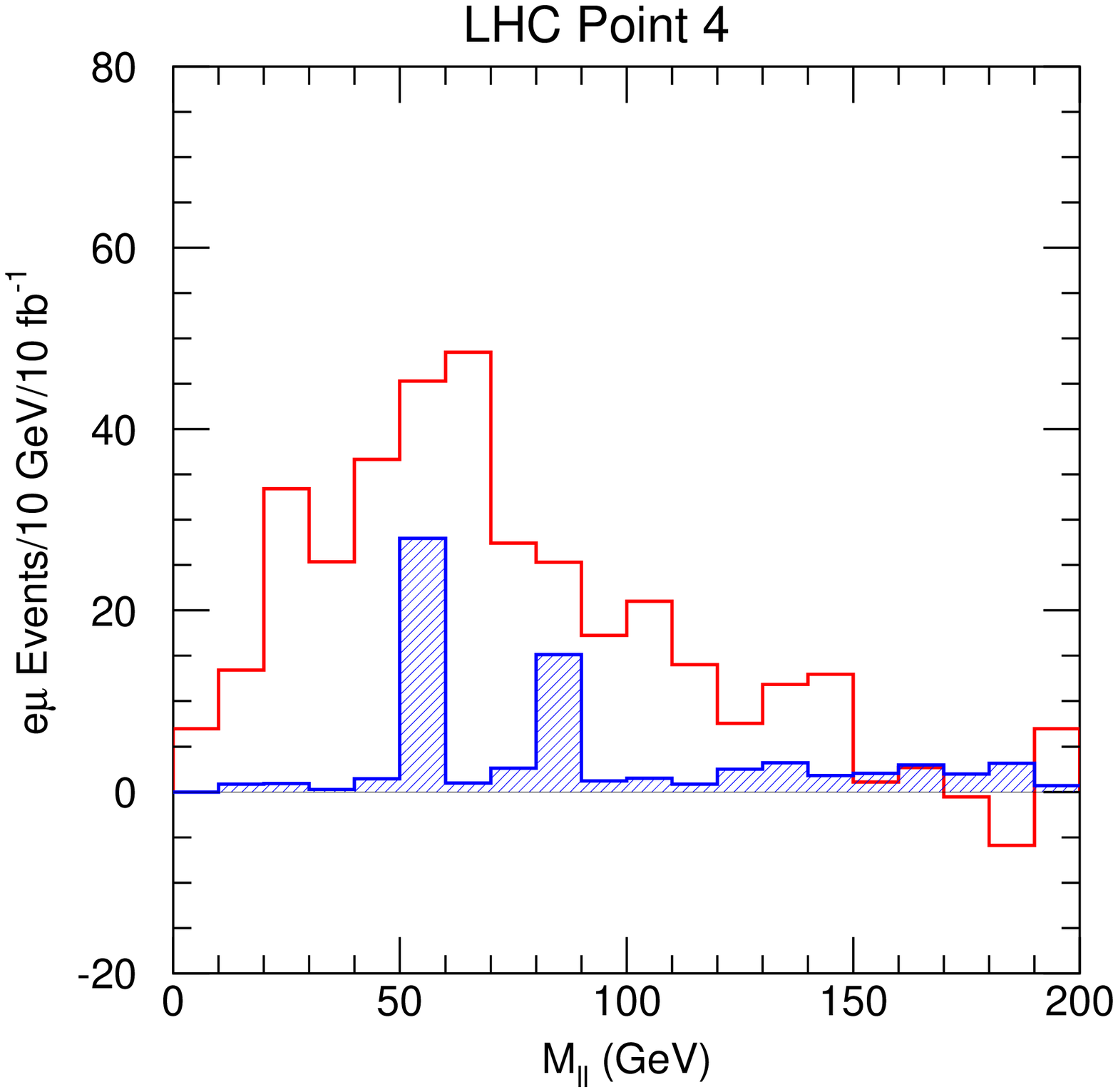}
\caption{Difference of the $M_{\ell^+\ell^-}$ distribution for
opposite-sign, opposite-flavor and same-sign, opposite-flavor
dileptons for the Point~4 signal (open histogram) and the Standard Model
background (shaded histogram).\label{c4_mllemudif}}
\end{figure}

	There is a large background to $e^\pm \mu^\mp$ from $t \bar t$
production. To suppress this it is necessary to raise the cut on
$\Meff$ from $800\,\GeV$ to $1000\,\GeV$. The like-sign and
opposite-sign $e\mu$ mass distributions with this cut and the other
cuts described in the previous subsection are shown in
Figure~\ref{c4_mllemu} together with the Standard Model backgrounds.
The difference of the like-sign and opposite-sign distributions is
shown in Figure~\ref{c4_mllemudif}. The Standard Model background
shows statistical fluctuations but is fairly small after
these cuts. 

	For Point~4 the endpoint of the $e^\pm\mu^\mp$ mass
distribution is determined by $\tchi_4^0 \ra \tchi_1^\pm W^\mp \ra
e^\pm \mu^\mp \lsp$ and is $220.6\,\GeV$. This is consistent with
Figure~\ref{c4_mllemudif}. Of course other models might lead to the
dominance of this decay by other modes. Because there are two missing
neutrinos in addition to the $\lsp$,
 there is no sharp edge at the kinematic limit, so the
endpoint can be determined only roughly. However, the total number of
events in Figure~\ref{c4_mllemudif} can be measured to a few percent; it
provides a measure of another combination of branching ratios.

\newpage

\newpage

\newpage

\newpage

\newpage
\section{LHC Point 5: $m_0=100\,\GeV$, $\mhalf=300\,\GeV$, $\tan\beta=2.1$}
\label{sec:lhc5}

	LHC Point~5 has a gluino with mass 767~GeV and light squarks
with masses of 662--690~GeV, so $\tg \ra \tq \bar q$ dominates. It has
$M(\tchi_2^0) = 232.6\,\GeV$, $M(\lsp) = 121.7\,\GeV$, and $M(h) =
104.15\,\GeV$, so $\tchi_2^0 \ra \lsp h$ is kinematically allowed. It
also has light right-handed sleptons, $M(\tilde\ell_R) = 157.2\,\GeV$,
so that $\tchi_2^0 \ra \tilde\ell_R^\pm \ell^\mp \ra \lsp\ell^+\ell^-$
also has a large branching ratio. This point was chosen so that the
$\lsp$ provides the correct amount of cold dark matter for cosmology;
this generally requires relatively light sleptons\cite{BB}.

\subsection{Selection of $h \ra b \bar b$ and Measurement of $M(\tilde
u_L)-M(\lsp)$} 

	For LHC Point~5 the decay chain $\tilde\chi_2^0 \ra \lsp h$,
$h \ra b \bar b$ has a large branching ratio, as is typical if this
decay is kinematically allowed. The decay $h \ra b \bar b$ thus
provides a handle for identifying events containing
$\tilde\chi_2^0$'s\cite{ERW}. Furthermore, the gluino is heavier than
the squarks and so decays into them. The strategy for this analysis is
to select events in which one squark decays via
$$
\tilde q \ra \tilde\chi_2^0 q,\ \tilde\chi_2^0 \ra \lsp h,\
h \ra b \bar b\,,
$$
and the other via
$$
\tilde q \ra \lsp q\,,
$$
giving two $b$ jets and exactly two additional hard jets.

	ISAJET~7.22\cite{ISAJET} was used to generate a sample of 100K
events for Point~5, corresponding to about $5.6\,{\rm fb}^{-1}$ so the
signal statistics shown in this section roughly correspond to the
actual statistics expected in 1 year at low luminosity.  The
background samples generally represent a small fraction of an LHC
year. The detector response was simulated using the toy calorimeter
described above.  Jets were found using a fixed cone algorithm with $R
= 0.4$. The following cuts were imposed:
\begin{itemize}
\item	$\etmiss > 100\,\GeV$;
\item	$\ge4$ jets with $p_T > 50\,\GeV$ and  $p_{T,1} > 100\,\GeV$;
\item	Transverse sphericity $S_T > 0.2$;
\item	$\Meff > 800\,\GeV$;
\item	$\etmiss > 0.2 \Meff$.
\end{itemize}
As before, jets were tagged as $b$'s if they contained a $B$ hadron
with $p_T > 5\,\GeV$ and $\eta < 2$, and a tagging efficiency of 60\%
per $b$ was included. 

	Figure~\ref{c5_mbb} shows the resulting $b \bar b$ mass
distributions for the signal and the sum of all Standard Model
backgrounds with $p_{T,b} > 25\,\GeV$ together with a Gaussian plus
quadratic fit to the signal.  The mistagging background is
comparable to the real background shown. The energy calorimeter scale for $b$-jets
was recalibrated to bring the Higgs mass peak to it's correct value
which will be measured ultimately via the decay to $\gamma\gamma$.
The correction is about 8\%. Using a larger cone, $R = 0.7$, gives an
uncorrected peak which is closer to the true mass but wider. Note that
for Point~5 the light Higgs could be discovered in this channel with
much less integrated luminosity than is needed to observe $h \ra
\gamma\gamma$; the latter would provide a better mass measurement,
$\Delta M_h < 1\,\GeV$. 

	Events were then required to have exactly one $b \bar b$ pair
with invariant mass within $\pm 1.5 \sigma$ ($\sim 19$ GeV)
of the Higgs peak and exactly
two additional jets with $p_T > 75\,\GeV$. The invariant mass of each
jet with the $b \bar b$ pair was calculated.  For the desired decay
chain, one of these two must come from the decay of a single squark,
so the smaller of them must be less than the kinematic limit for the
decay chain $\tq \ra \tchi_2^0 q \ra \lsp h q$, namely
$$
(M_{hq}^{\rm max})^2 = M_h^2 
+ \left(M_{\tq}^2 - M_{\tchi_2^0}^2\right) 
\left[{M_{\tchi_2^0}^2 + M_h^2 - M_{\lsp}^2 + 
\sqrt{(M_{\tchi_2^0}^2 - M_h^2 - M_{\lsp}^2)^2 -4M_h^2
M_{\lsp}^2}} \over 2 M_{\tchi_2^0}^2 \right]\,.
$$
Using the average of the $u_L$ and $d_L$ masses gives $M_{hq}^{\rm
max} = 506\,\GeV$. The smaller of the two $b \bar b j$ masses is
plotted in Figure~\ref{c5_mbbj} for the signal and for the sum of all
backgrounds and shows an edge near the expected value. The Standard Model
background shows fluctuations from the limited Monte Carlo statistics
but seems to be small near the edge, at least for the idealized
detector considered here. There is some background from the SUSY
events above the edge, presumably from other decay modes and/or
initial state radiation. 

	A detailed understanding of the shape of this edge and its
relation to the masses involved requires more study.  Based on the
statistics in Figure~\ref{c5_mbbj}, it seems likely that one could
determine the end point of the spectrum to $\sim 40\,\GeV$ in one year
and to half of that in three years at low luminosity.

\subsection{Selection of $W \ra q \bar q$ and Measurement of $M(\tilde
u_L)-M(\lsp)$} 

	Point~5 also has a large combined branching ratio for one
gluino to decay via
$$
\tilde g \ra \tilde q_L \bar q,\ \tilde q_L \ra \tilde\chi_1^\pm q,\
\tilde\chi_1^\pm \ra \lsp W^\pm,\ 
W^\pm \ra q \bar q\,,
$$
and the other via
$$
\tilde g \ra \tilde q_R q,\ \tilde q_R \ra \lsp q\,,
$$
giving two hard jets and two softer jets from the $W$. The branching
ratio for $\tilde q_L \ra \lsp q$ is small for Point~5, so the
contributions from $\tilde g \ra \tilde q_L \bar q$ and from $\tilde
q_L \tilde q_L$ pair production are suppressed.

	The same signal sample was used as in the previous subsection.
The combinatorial background for this decay chain is much larger than
for the previous one, so harder cuts are needed:
\begin{itemize}
\item	$\etmiss > 100\,\GeV$;
\item	$\ge4$ jets with $p_{T1,2} > 200\,\GeV$, $p_{T3,4} > 
50\,\GeV$, and $\eta_{3,4} < 2$;
\item	Transverse sphericity $S_T > 0.2$;
\item	$\Meff > 800\,\GeV$;
\item	$\etmiss > 0.2 \Meff$.
\end{itemize} 
The same $b$-tagging algorithm was applied to tag the third and fourth
jets as not being $b$ jets. In practice one would measure the $b$-jet
distributions and subtract them.  

\begin{figure}[h]
\dofig{3.20in}{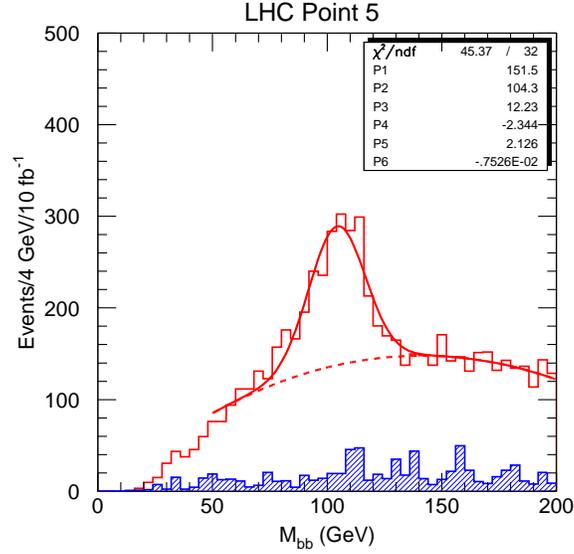}
\caption{$M(b \bar b)$ for pairs of $b$ jets for the Point~5 signal
(open histogram) and for the sum of all backgrounds (shaded histogram)
after cuts described in the text. The smooth curve is a Gaussian plus
quadratic fit to the signal. The light Higgs mass is
$104.15\,\GeV$.\label{c5_mbb}}
\end{figure}

	The mass distribution $M_{34}$ of the third and fourth highest
$p_T$ jets with these cuts is shown in Figure~\ref{c5_mjj} for the
signal and the sum of all backgrounds. A peak is seen a bit below the
$W$ mass with a fitted width  smaller than that for the
$h$ in Figure~\ref{c5_mbb}; note that the $W$ natural width has been
neglected in the simulation of the decays.\footnote{The Higgs width is
much smaller than the $W$ width for Higgs masses relevant to these
analyses.} The Standard Model background is more significant here than
for the $h \ra b \bar b$ channel. Events from this peak can be
combined with another jet as was done for $h \ra b
\bar b$, providing another determination of the squark mass,
Figure~\ref{c5_mjjj}, with an error similar to the previous one.
Figure~\ref{c5_mjj} also provides a starting point for measuring the $W$
production rate in SUSY events.  Knowing this rate is essential
when searching for excess leptons from other sources such as gaugino decays.

\subsection{Selection of $\tilde\chi_2^0 \ra \tilde\ell \ell \ra \lsp
\ell\ell$} 

	Point~5 has relatively light sleptons, as is generically
necessary if the $\lsp$ is to provide acceptable cold dark
matter\cite{BB}, since $\lsp-\lsp$ annihilation in the early universe
proceeds via slepton exchange and the slepton mass must therefore be
small enough to make this rate sufficiently large. Hence the two-body
decay
$$
\tilde\chi_2^0 \ra \tilde\ell_R^\pm \ell^\mp \ra \lsp \ell^+\ell^-
$$
is kinematically allowed and competes with the $\tilde\chi_2^0 \ra
\lsp h$ decay, producing opposite-sign, like-flavor dileptons. This
source of sleptons is much larger than that from direct production and
their discovery is much easier than at Point 3 despite the fact that
the total sparticle production rate is much larger at that point.  The
largest Standard Model background is $t \bar t$. To suppress this and
other Standard Model backgrounds the following cuts were made on the
same signal and Standard Model background samples used previously:
\begin{itemize}
\item	$\Meff > 800\,\GeV$;
\item	$\etmiss > 0.2\Meff$;
\item	$\ge 1$ $R=0.4$ jet with $p_{T,1} > 100\,\GeV$;
\item	$\ell^+\ell^-$ pair with $p_{T,\ell}> 10\,\GeV$, $\eta_\ell <
2.5$;
\item	$\ell$ isolation cut: $E_T < 10\,\GeV$ in $R=0.2$;
\item	Transverse sphericity $S_T > 0.2$.
\end{itemize}
With these cuts very little Standard Model background survives, and the
$M_{\ell\ell}$ mass distribution shown in Figure~\ref{c5_mll} has an
edge near the kinematic limit for this decay sequence, namely
$$
M_{\ell\ell}^{\rm max} = M_{\tilde\chi_2^0} 
\sqrt{1-{M_{\tilde\ell}^2 \over M_{\tilde\chi_2^0}^2}}
\sqrt{1-{M_{\lsp}^2 \over M_{\tilde\ell}^2}} \approx 108.6\,\GeV\,,
$$
Observing both $h \ra b \bar b$ with $M_h>M_Z$ and an $\ell^+\ell^-$
continuum with $M_{\ell\ell}>M_Z$ would certainly suggest, and perhaps
establish, the existence of light sleptons. 

\begin{table}[t]
\caption{Comparison of masses relevant to dilepton spectrum for
Point~5 and for modified point with $m_0 = 120\,\GeV$.\label{masses}}
\begin{center}
\begin{tabular}{ccc}
\hline\hline
Mass		& Point~5	& With $m_0 = 120\,\GeV$ \\
\hline
$\tg$		& 767.1~GeV	& 767.2~GeV \\
$\tchi_2^0$	& 231.2~GeV	& 231.4~GeV \\
$\tilde\ell_R$	& 157.2~GeV	& 170.6~GeV \\
$\lsp$		& 121.3~GeV	& 121.4~GeV \\
\hline\hline
\end{tabular}
\end{center}
\end{table}

\begin{figure}[h]
\dofig{3.20in}{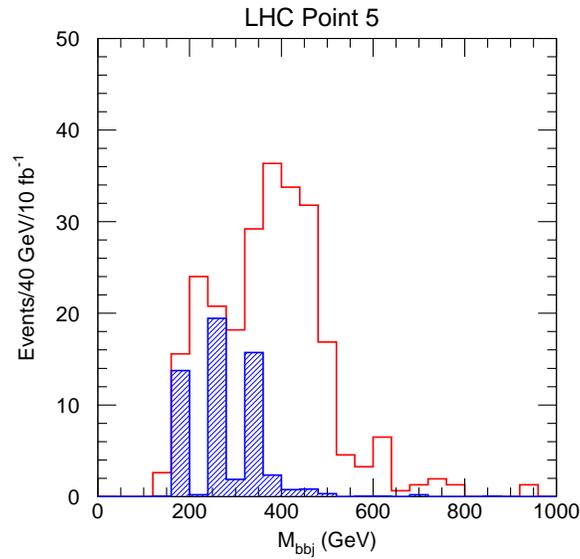}
\caption{The smaller of the two $b \bar b j$ masses for the signal and
background events with $73 < M(b \bar b) < 111\,\GeV$ in
Figure~\protect\ref{c5_mbb} and exactly two additional jets $j$ with
$p_T > 75\,\GeV$. The endpoint of
this distribution should be approximately $M_{hq}^{\rm max} =
506\,\GeV$.\label{c5_mbbj}} 
\end{figure}

\begin{figure}[h]
\dofig{3.20in}{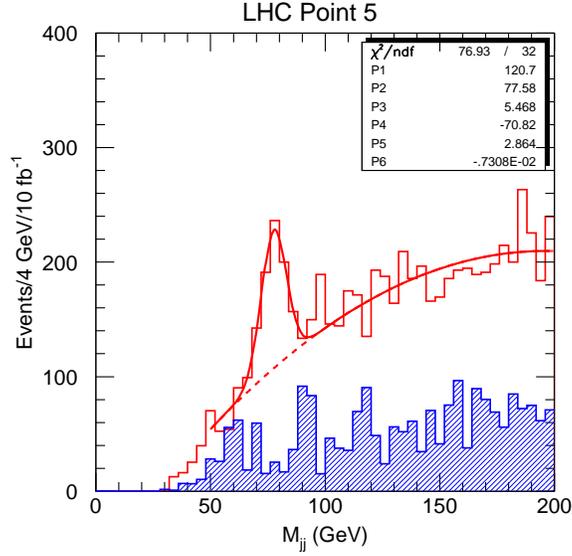}
\caption{$M_{34}$ for non-$b$ jets in events with two $200\,\GeV$
jets and two $50\,\GeV$ jets for the Point~5 signal (open histogram)
and the sum of all backgrounds (shaded histogram).\label{c5_mjj}}
\end{figure}

\begin{figure}[h]
\dofig{3.20in}{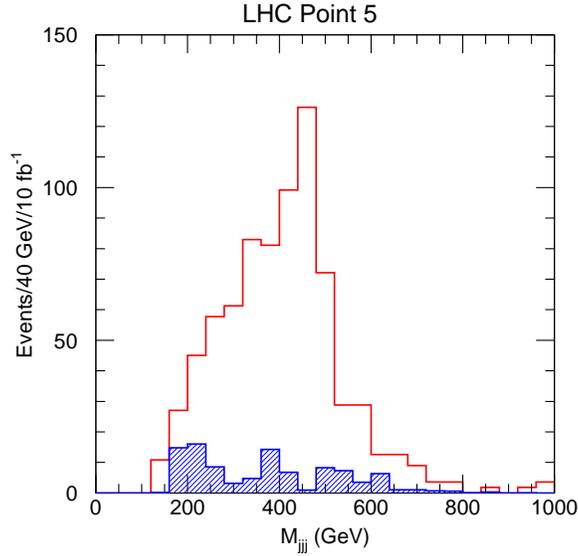}
\caption{The smaller of the two $q \bar q j$ masses for signal and
background (shaded) events with $71 < M(b \bar b) < 87\,\GeV$ in the previous
figure and with exactly two additional jets $j$ with $p_T > 75\,\GeV$.
The endpoint of this distribution should be approximately the mass
difference between the squark and the $\tilde \chi_1^0$, about
$565\,\GeV$.\label{c5_mjjj}} 
\end{figure}

\begin{figure}[h]
\dofig{3.20in}{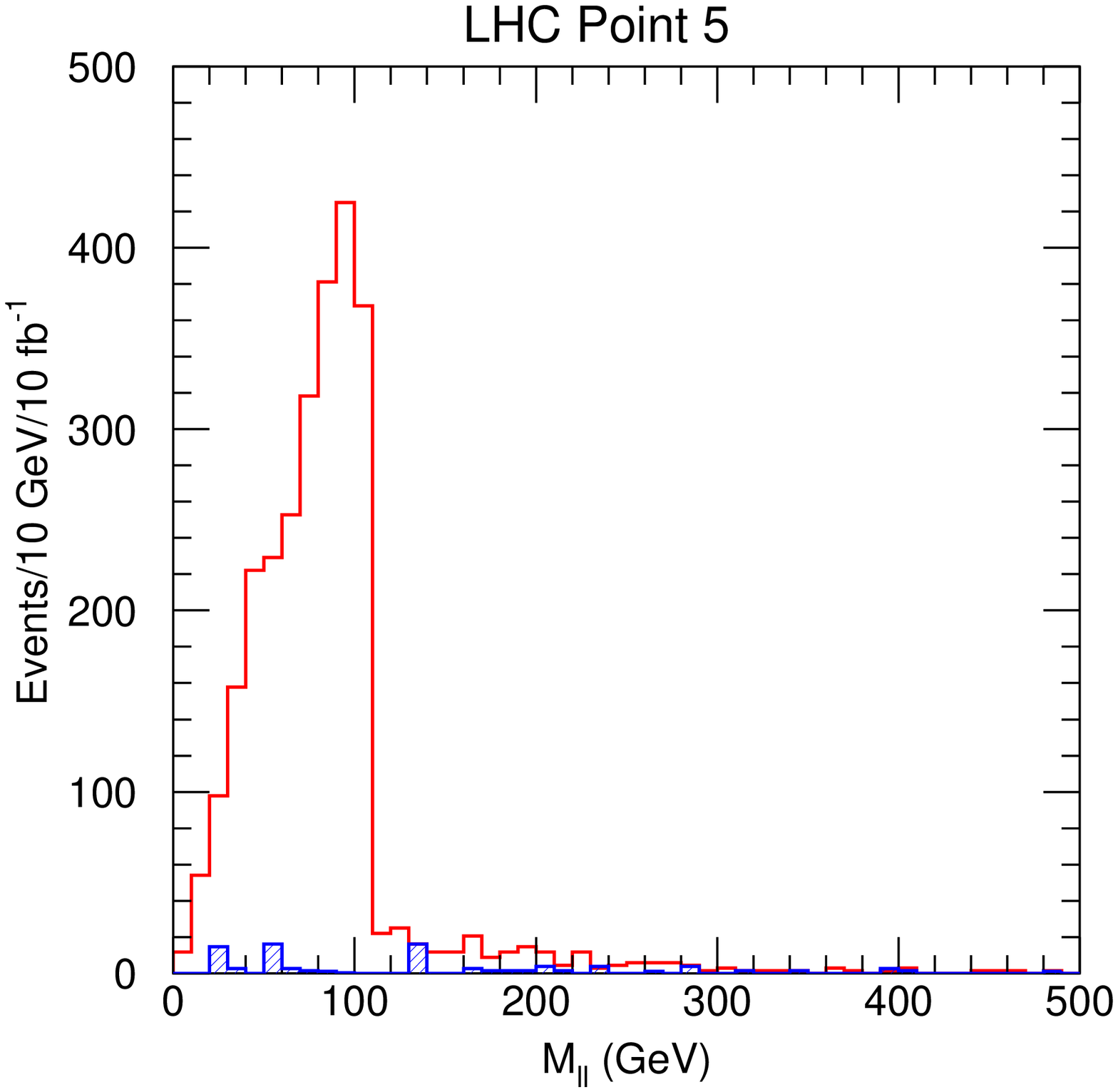}
\caption{$M_{\ell\ell}$ for the Point~5 signal (open histogram) and
the sum of all backgrounds (shaded histogram).\label{c5_mll}}
\end{figure}

\begin{figure}[h]
\dofig{3.20in}{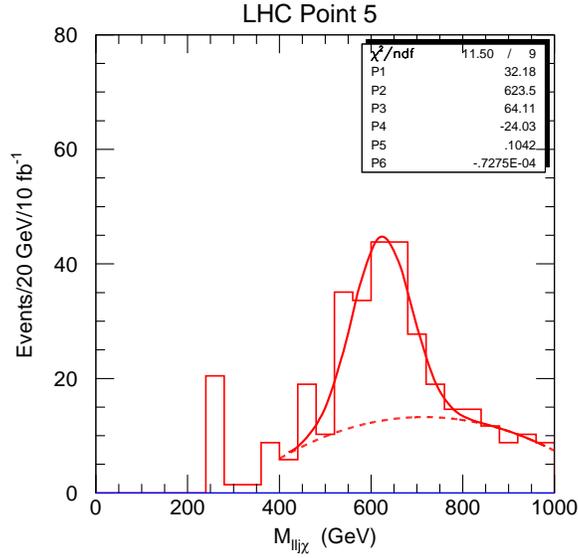}
\caption{$M_{\ell\ell j\lsp}$ for events with $86 < M_{\ell\ell} <
109\,\GeV$ using $\vec p_{\lsp} = M_{\lsp} / M_{\ell\ell} 
\vec p_{\ell\ell}$ for the Point~5 signal (open histogram) and the
Standard Model background (shaded histogram).\label{c5_mlljh}}
\end{figure}

	If $M_{\ell\ell}$ is near its kinematic limit, then the
velocity difference between the $\ell^+\ell^-$ pair and the $\lsp$ is
minimized in the rest frame of $\tilde\chi_2^0$. Having both leptons
hard requires $M_{\tilde\ell}/ M_{\tilde\chi_2^0}^2 \sim M_{\lsp} /
M_{\tilde\ell}$. Assuming this and $M_{\tilde\chi_2^0} = 2 M_{\lsp}$
implies that the endpoint in Figure~\ref{c5_mll} is equal to the
$\lsp$ mass. An improved estimate could be made by detailed fitting of
all the kinematic distributions.  Events were selected with
$M_{\ell\ell}^{\rm max} -10\,\GeV < M_{\ell\ell} < M_{\ell\ell}^{\rm
max}$, and the $\lsp$ momentum was calculated using this crude $\lsp$
mass and
$$
\vec p_{\lsp} = (M_{\lsp} / M_{\ell\ell})\,\vec p_{\ell\ell}\,.
$$
The invariant mass $M_{\ell\ell j\lsp}$ of the $\ell^+\ell^-$, the
highest $p_T$ jet, and the $\lsp$ was then calculated and is shown in
Figure~\ref{c5_mlljh}.  A broad peak is seen near the light squark
masses, 660--$688\,\GeV$.
This peak contains complementary information to that obtained from 
Figure~\ref{c5_mbbj} and more information about the masses could be
obtained by performing a combined fit.

\begin{figure}[h]
\dofig{3.20in}{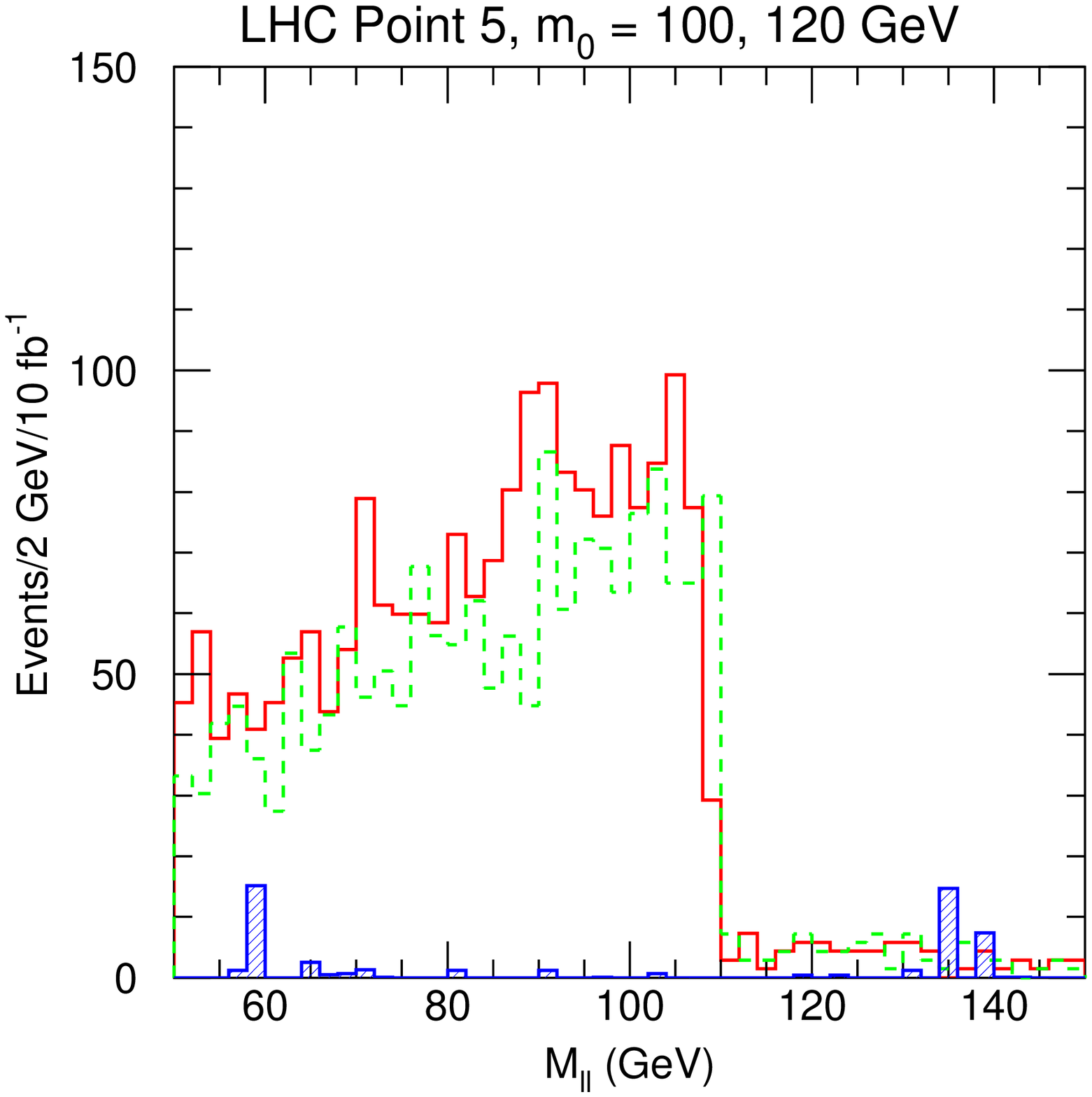}
\caption{$M(\ell\ell)$ for Point~5 (solid curve) and for a modified
point with $m_0 =120\,\GeV$ (dashed curve). The shaded histogram is
the sum of all Standard Model backgrounds.\label{c5p20_mll}} 
\end{figure}

\begin{figure}[h]
\dofig{3.20in}{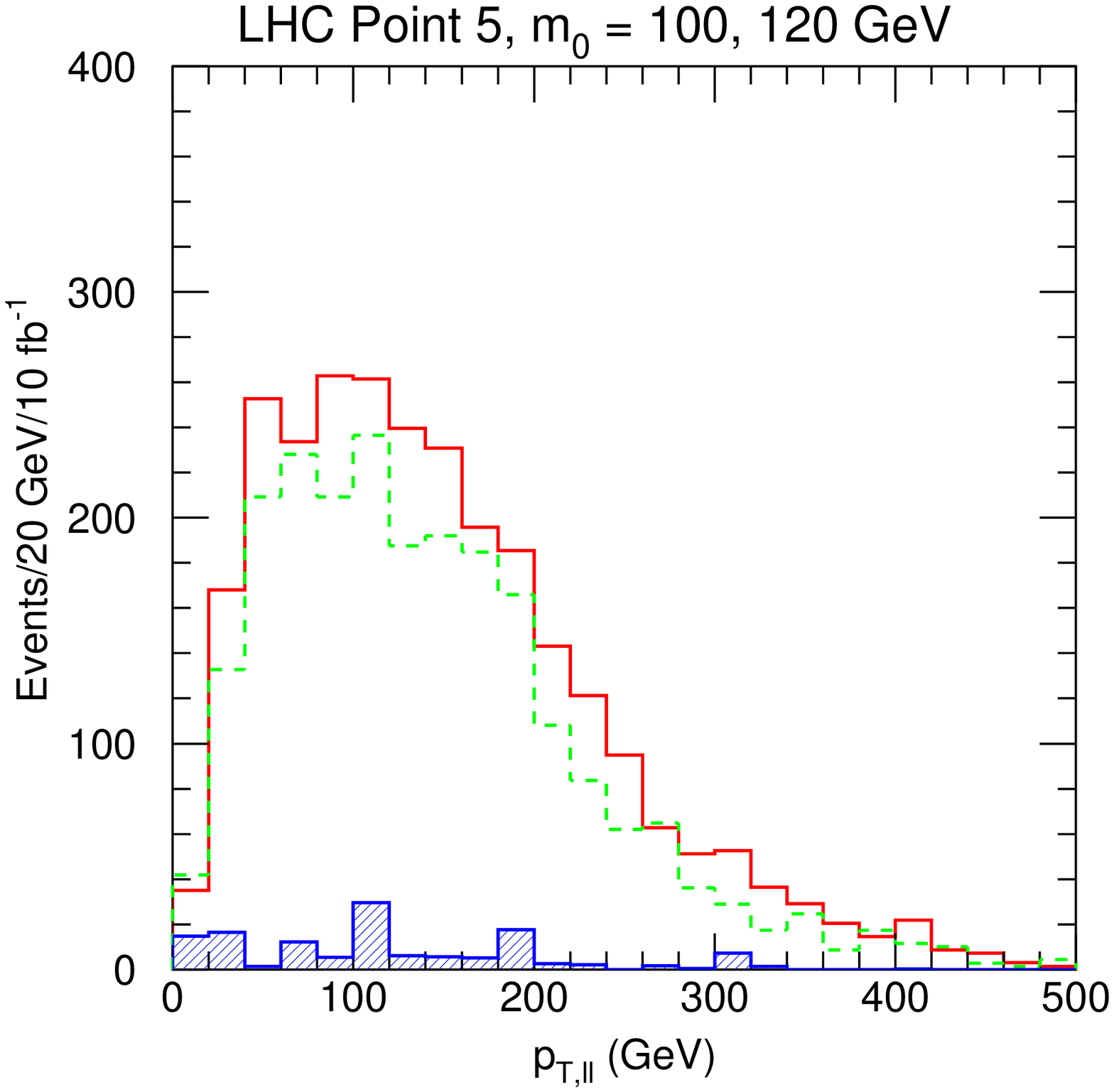}
\caption{$p_T(\ell\ell)$ for Point~5 (solid curve) and for a modified
point with $m_0 =120\,\GeV$ (dashed curve). The shaded histogram is
the sum of all Standard Model
 backgrounds.\label{c5p20_ptll}}
\end{figure}

	A detailed analysis of the dilepton mass spectrum would
require varying $M(\tchi_2^0)$, $M(\tilde\ell_R)$, $M(\lsp)$, and the
$p_T(\tchi_2^0)$ distribution and fitting the distributions of
$M(\ell\ell)$, $p_T(\ell\ell)$, $p_T(h)$, and
$p_T(\ell_2)/p_T(\ell_1)$. Rather than do this, a sample of 50K events
with $m_0 = 120\,\GeV$ but otherwise the same parameters was
generated. A comparison of some of the relevant masses is shown in
Table~\ref{masses}; the slepton mass changes by $13\,\GeV$, and the
rest are essentially identical. The $M(\ell\ell)$ mass distribution
near the edge is shown in Figure~\ref{c5p20_mll}.  There is a shift in
the location of the edge by about $2\,\GeV$, which should be
observable.  There is not much change in the $p_T$ distribution of the
$\ell\ell$ pair, Figure~\ref{c5p20_ptll}. The most sensitive
distribution is $p_T(\ell_2)/p_T(\ell_1)$, where by definition
$p_T(\ell_2) < p_T(\ell_1)$. There is a clear change in the shape, as
one would expect. For fixed values of the other parameters, one ought
to be able to determine $m_0$ to $\sim 5\,\GeV$, 
although careful subtraction of Standard Model backgrounds will be
necessary.

\subsection{Top Production in SUSY Events}

\label{subsec:top}
Gluino decays at LHC Point~5 have a sizable branching ratio to $\tilde
t \bar t$. We have attempted to isolate such a sample by searching for
 top decays. In order to reduce the background from Standard
Model top production, the cut on $\Meff$ has been raised to $\Meff >
1000\,\GeV$ for this analysis. The event selection is as follows:
\begin{itemize}
\item	$\etmiss > 100\,\GeV$;
\item	$\ge4$ jets with $p_{T1} > 100\,\GeV$, $p_{T2,3,4} > 
50\,\GeV$;
\item	Transverse sphericity $S_T > 0.2$;
\item	$\Meff > 1000\,\GeV$;
\item	$\etmiss > 0.2 \Meff$.
\item  Exactly two b-jets with $p_T > 25\,\GeV$. 
\end{itemize} 
If the invariant mass of the two-$b$
system is within $\pm 20$ GeV of the Higgs mass, the event is
rejected. To search for $W \rightarrow q\bar{q}$, we calculate the
invariant mass of all dijet combinations with both jets having $p_T >
50$ GeV and with neither jet being tagged as a b. Combinations with
$|M_{q\bar{q}}-M_W| < 10 \,\GeV$ mass are considered W candidates.
Combinations where $20 \,\GeV < |M_{q\bar{q}}-M_W| < 30\,\GeV$ are
used to model the shape of the background; we will refer to these
events as the $W$ sideband region. The $M_{q\bar{q}}$ distribution is
shown in Figure~\ref{c5-W-for-top} where a clear $W$ peak is visible.
The average jet multiplicity in these events is $~\sim 9$ and hence
there is a large combinatorial background in this figure.

Dijets in the $W$ mass and $W$ sideband regions are then combined with
each of the two $b$ jets, giving two combinations for each $W$.
Results of this reconstruction for the $W$ signal and sideband regions
are shown in Figure~\ref{c5-marjietopu}.  Subtraction of the sideband
distribution from the signal yields Figure~\ref{c5-marjietop}. A clear
top signal is seen. There is a small background of Standard Model
production of top quarks, shown as the hatched histogram on this
figure. The dominant decay chain giving rise to this signal is
\begin{eqnarray*}
\tilde{g} &\to& \tilde{t}t\\
\tilde{t} &\to& \tchi_1^+b\\
\tchi_1^+ &\to& \lsp W 
\end{eqnarray*}
The kinematics of the decay $\tilde{t} \to \tchi_1^+ b$, $\tchi_1^+
\to \lsp W$ restricts the $Wb$ invariant mass distribution to be
between 203 and 356~GeV. There is some evidence in
Figure~\ref{c5-marjietop} for an excess of events in this region.
Unfortunately, other possible combinations and decays such as
$\tilde{g}\to \tilde{b}b$, $\tilde{b}\to \tchi_1^+ t$, $\tchi_1^+ \to
\lsp W$, which has a smaller combined branching fraction, make the
exact interpretation difficult.
Thus, when discussing the determination of SUSY parameters in
Section~\ref{sec:scan-5}, we assume that top decays in SUSY events
can be seen at Point 5 (indicating that the $\tilde{g} \ra \tilde{t}t$ 
channel is open), but we will not assume that the
$\tilde{t}$ mass can be determined.

\newpage

\newpage

\newpage

\newpage

\newpage

\begin{figure}[h]
\dofig{3.20in}{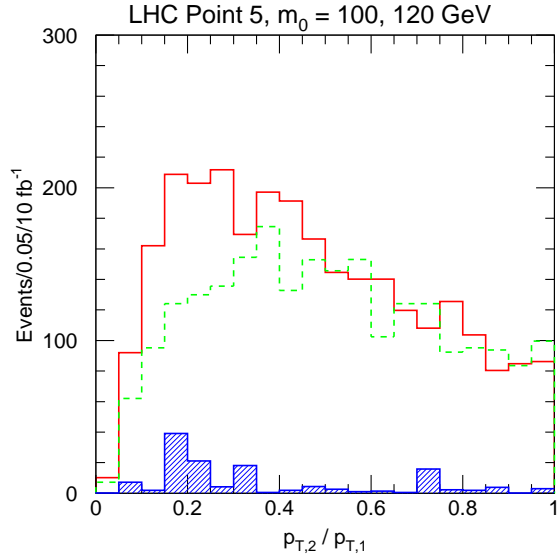}
\caption{$p_T(\ell_2)/p_T(\ell_1)$ for Point~5 (solid curve) and for a
modified point with $m_0 =120\,\GeV$ (dashed curve).
The shaded histogram shows the sum of all backgrounds.
\label{c5p20_pt2pt1}}
\end{figure}

\begin{figure}[h]
\dofig{3.20in}{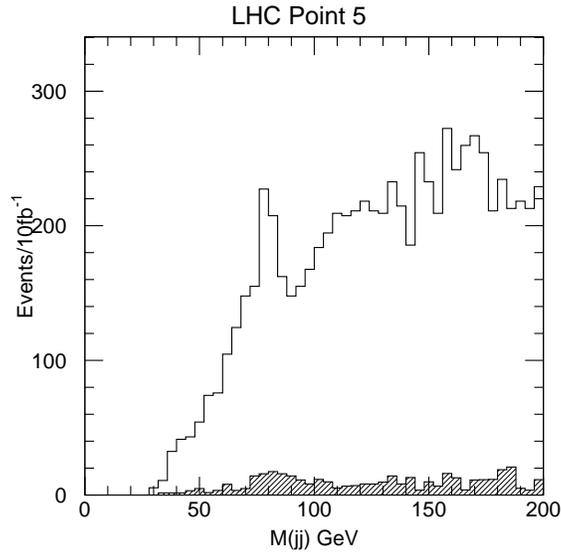}
\caption{The invariant mass distribution used to search for
$W\rightarrow q\bar{q}$ decays in the Point~5 top reconstruction.
The shaded histogram shows the sum of all backgrounds.
\label{c5-W-for-top}}
\end{figure}   

\newpage

\begin{figure}[h]
\dofig{3.20in}{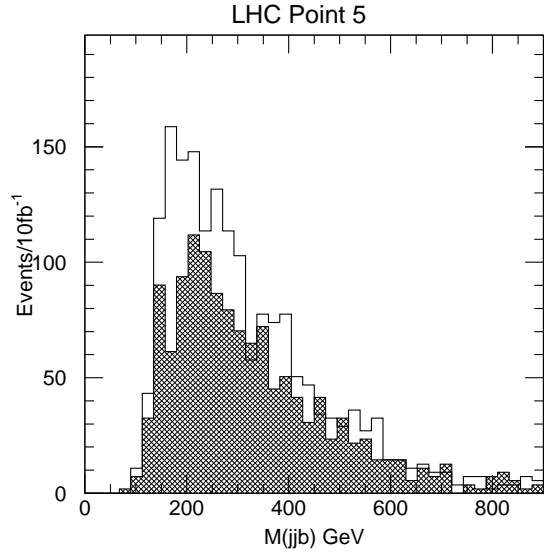}
\caption{The invariant mass distribution of $q\bar{q}b$
candidates for combinations where the $q\bar{q}$ pair are in the
$W$ mass region (solid) and $W$ sideband region (dashed). The $b$-jet
energies have {\bf not} been recalibrated and a tagging efficiency of
60\% per b included. Note that each event appears
twice.\label{c5-marjietopu}}
\end{figure}   

\begin{figure}[h]
\dofig{3.20in}{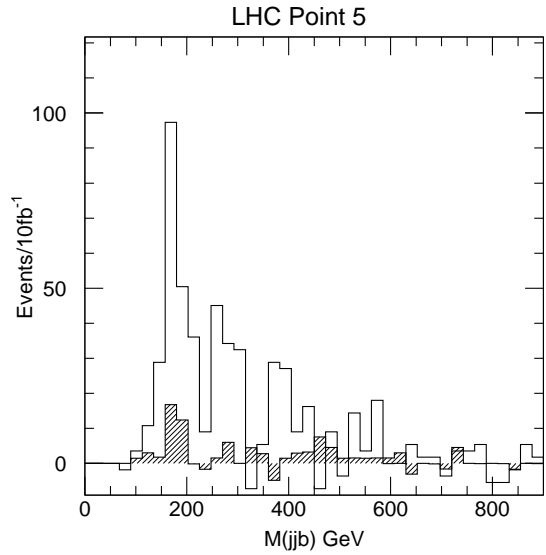}
\caption{The sideband subtracted
invariant mass distribution of $Wb$ candidates.  The hatched distribution
shows the contribution from Standard Model production of top quarks.
\label{c5-marjietop}}
\end{figure}   

\newpage
\section{LHC Points 1, 2:  $m_0=\mhalf=400\,\GeV$, $\tan\beta=2,10$}
\label{sec:lhc1}

	LHC Points 1 and 2 have gluino masses of about 1~TeV and
squark masses about 50~GeV lighter. These are close to the upper limit of
the expected range if SUSY is to be relevant to electroweak symmetry breaking.
 The cross sections are quite small, so more than
$10\,\fb^{-1}$ is needed for precision studies. Only a few results
will be presented here.

\subsection{Selection of $h \ra b \bar b$ and Measurement of $M(\tilde
u_L)-M(\lsp)$} 

\begin{figure}[h]
\dofig{3.20in}{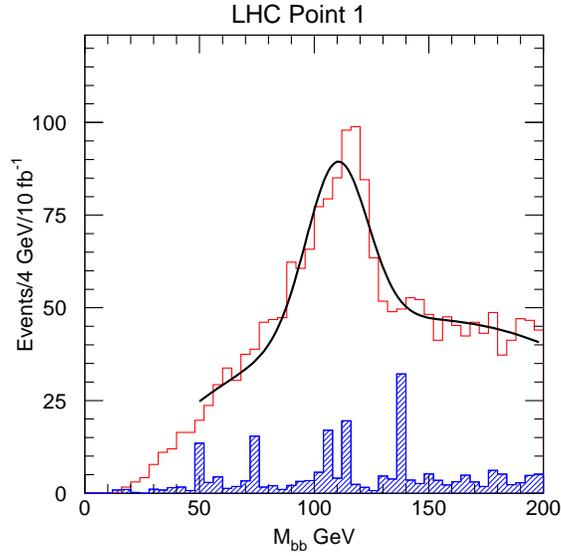}
\caption{$M(b \bar b)$ for pairs of $b$ jets for the LHC Point~1 signal
(open histogram) and for the sum of all backgrounds (shaded histogram)
after cuts described in the text. The smooth curve is a Gaussian plus
quadratic fit to the signal. The light Higgs mass is $111.2\,\GeV$.
\label{c1_mbb}}
\end{figure}

\begin{figure}[h]
\dofig{3.20in}{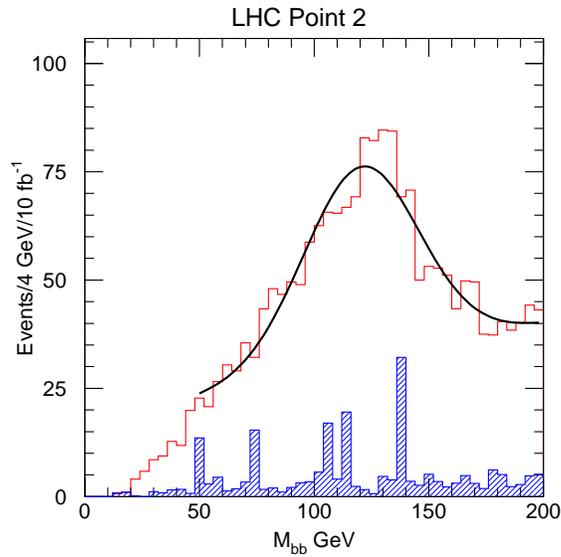}
\caption{Same as Figure~\protect\ref{c1_mbb} for LHC Point~2. The light
Higgs mass is $125.1\,\GeV$.\label{c2_mbb}}
\end{figure}

	This analysis is very similar to the one for LHC Point~5, but
since the signal cross sections are smaller, harder cuts are needed:
\begin{itemize}
\item	$\etmiss > 100\,\GeV$;
\item	$\ge4$ jets with $p_T > 50\,\GeV$, $p_{T,1} > 250\,\GeV$
$p_{T,2}>150\,\GeV$;
\item	Transverse sphericity $S_T > 0.2$;
\item	$\Meff > 1000\,\GeV$;
\item	$\etmiss > 0.2 \Meff$.
\end{itemize}
Events with exactly two tagged $b$ jets were then selected.
Figures~\ref{c1_mbb} and \ref{c2_mbb} show the $b \bar b$ mass
distributions for Points 1 and 2 respectively.  The light Higgs would
be discovered in this mode at either point with $10\,\fb^{-1}$ or
less, although $h \ra \gamma\gamma$ would still be needed to provide
the most precise mass determination.

\begin{figure}[h]
\dofig{3.20in}{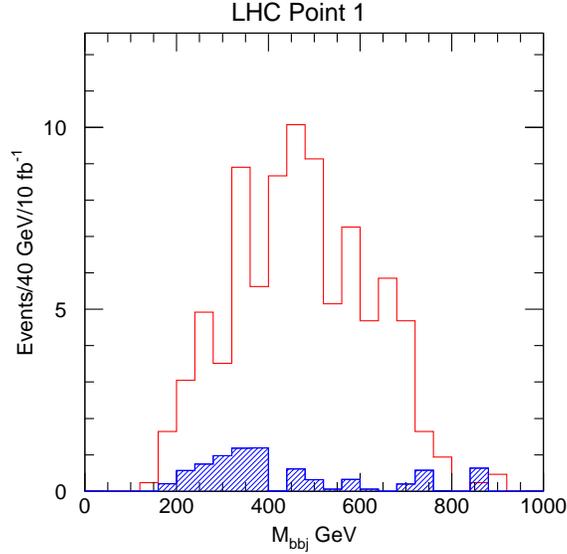}
\caption{The smaller of the two $b \bar b j$ masses for the signal and
background events with $M(b \bar b)$ within $2\sigma$ of the peak in
Figure~\protect\ref{c1_mbb} and exactly two additional jets $j$ with
$p_T > 100\,\GeV$. The endpoint of this distribution should be
approximately $M_{hq}^{\rm max} = 739\,\GeV$.\label{c1_mbbj}} 
\end{figure}

\begin{figure}[h]
\dofig{3.20in}{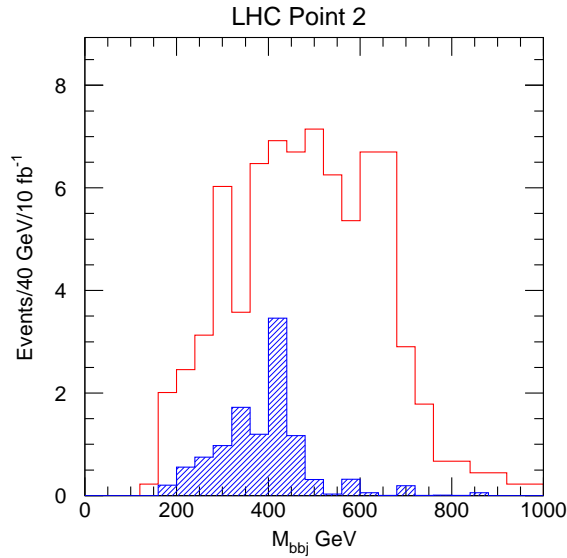}
\caption{Same as Figure~\protect\ref{c1_mbb} for LHC Point~2. The
endpoint should be $M_{hq}^{\rm max} = 751\,\GeV$.\label{c2_mbbj}}
\end{figure}

	Events passing these cuts and having a $b\bar b$ mass within
$2\sigma$ of the peak were then selected. These events were also
required to have two and only two additional jets with
$p_T>100\,\GeV$. This cut is quite inefficient, but attempts to
improve its efficiency produced more background. Each of the two jets
was combined with the $b\bar b$ pair. The smaller of the two masses
was selected for each event and plotted in Figures.\ref{c1_mbbj} and
\ref{c2_mbbj}. The endpoints for the decay sequence
$\tq_L \ra \tchi_2^0 q \ra \lsp b \bar b q$ are 739~GeV and 751~GeV
respectively. Measurement of this edge will be limited by statistics
to $\sim 50\,\GeV$ for $10\,\fb^{-1}$. 

\subsection{Selection of $\tchi_i \ra \tchi_j \ell^+\ell^-$}

	Sleptons are quite heavy for LHC Points 1 and 2, so there is
no edge in the dilepton mass spectrum as for Point~5. There is,
however, a $Z \ra \ell^+\ell^-$ signal that can be used to distinguish
these otherwise rather similar points.

\begin{figure}[h]
\dofig{3.20in}{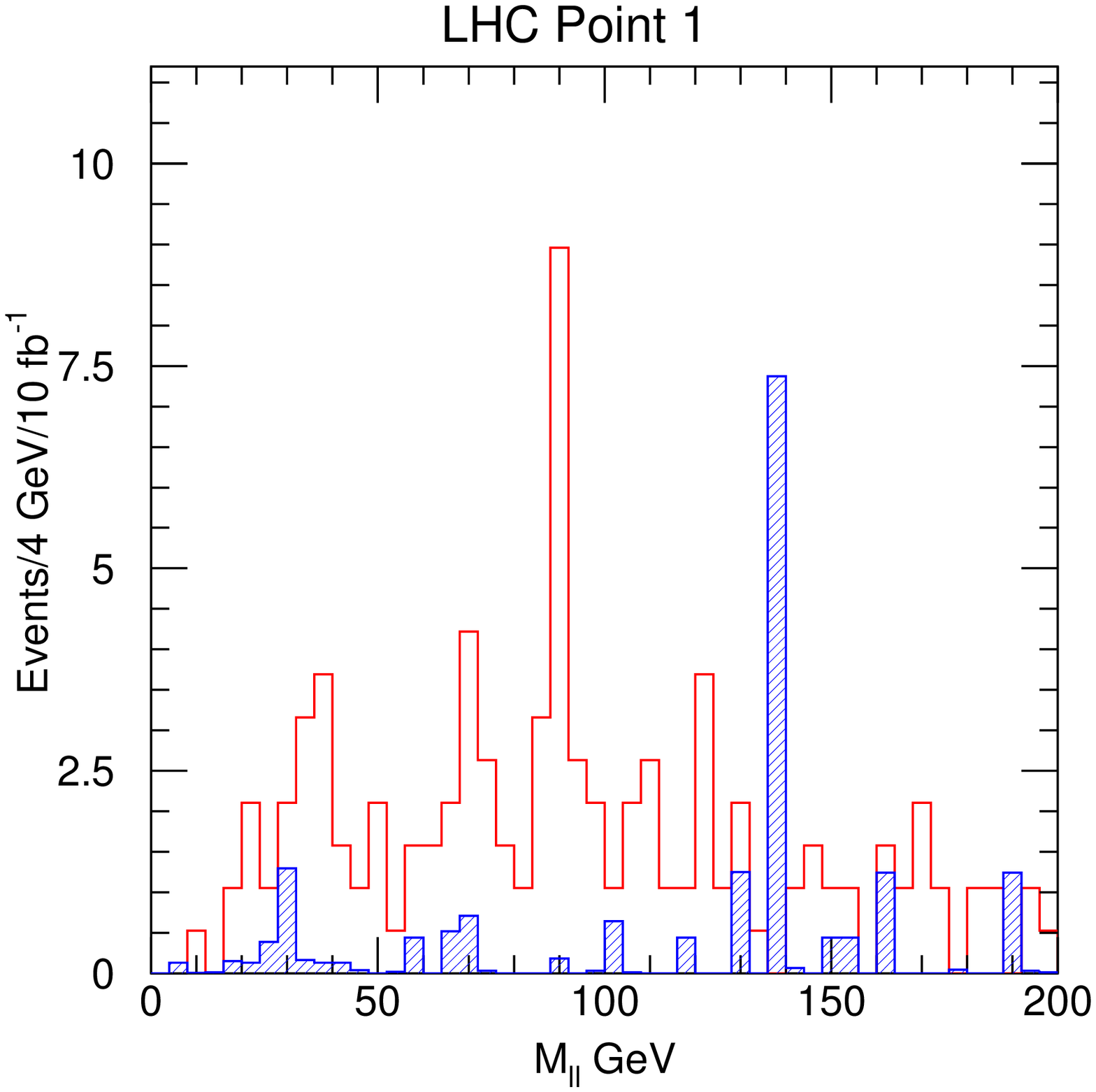}
\caption{$M_{\ell\ell}$ for the Point~1 signal (open histogram) and
the sum of all backgrounds (shaded histogram).\label{c1_mll}}
\end{figure}

\begin{figure}[h]
\dofig{3.20in}{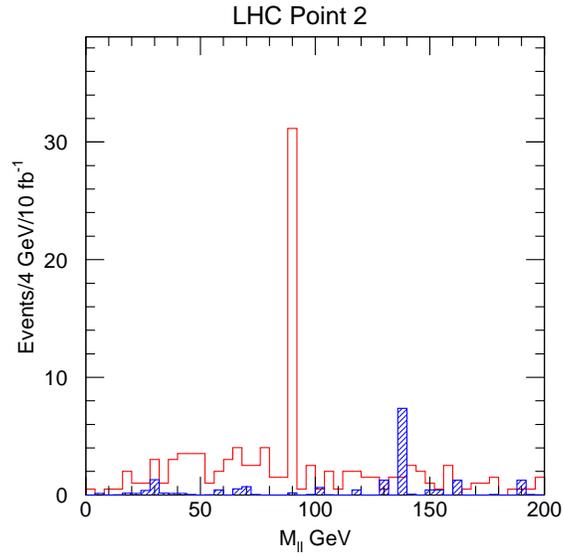}
\caption{$M_{\ell\ell}$ for the Point~2 signal (open histogram) and
the sum of all backgrounds (shaded histogram).\label{c2_mll}}
\end{figure}

	The basic selection cuts given at the beginning of the
previous subsection were applied. In addition, the events were
required to have two opposite-sign, same-flavor leptons with
$p_T>10\,\GeV$, $|\eta|<2.5$, and $E_T<10\,\GeV$ in a cone $R=0.2$.
The mass of the two highest $p_T$ such leptons is shown in
Figure~\ref{c1_mll} and Figure~\ref{c2_mll} for LHC Points 1 and 2
respectively. The $Z$ peak is questionable for Point~1 but rather
clear for Point~2, which has larger $\tan\beta$ and hence more mixing
of gauginos and Higgsinos. Even for Point~2, however, the statistical
error on the number of $Z$'s is $\sim15\%$. Clearly this is a
measurement that needs higher luminosity. 

	There are also observable signals in the like-sign and
in the opposite-sign, opposite-flavor dilepton channels. These would
certainly be useful in a global fit, but they do not seem to provide
the sort of precise measurements being considered in this paper.

\newpage

\newpage

\newpage

\newpage
\section{Determining SUSY parameters}
\label{sec:scan}

Once a number of quantities have been measured, we can attempt to
determine the particular SUSY model and the values of the parameters.
The strategy will be to attempt to perform a global fit to the model
parameters using all of the available data, much as the Standard Model
is tested using the $W$ and $Z$ masses and the many quantities
precisely measured by LEP/SLC. Such a fit is beyond the scope of our
work, and we adopt a simpler procedure. We assume that from
measurements of global parameters such as those discussed in
section~\ref{sec:meff} we know the approximate scale of the superpartner
masses and have some idea that we might be in a SUGRA model. The
object is then to determine the parameters of that model and check its
consistency. We must therefore determine the parameters $m_0$,
$m_{1/2}$, $A_0$, $\tan\beta$ and $\sgn\mu$. As we will see $A_0$ is
difficult to determine. Its value is given at the unification scale,
and the value that is relevant for the phenomenology is the one
evolved down to the electroweak scale. Many choices of $A_0$ evolve to
the same (fixed point) value, so there is reduced sensitivity to its value
at the unification scale. In addition it always appears scaled by the Yukawa
coupling for the relevant quark or lepton. Hence its effect on lepton masses and
on the quarks of the first two generations is very small. Its effect is
significant for top squarks and at large $\tan\beta$ for bottom squarks.
\footnote{The Higgs mass that we use is that determined from 1-loop calculations as
implemented in ISAJET\cite{ISAJET}. Two loop corrections to the Higgs mass,
implemented,
 for example, in SPTHYIA\cite{spythia} lower the mass by about 5\%.}

Our strategy for determining the parameters is as follows. We choose a
point randomly in parameter space and compute the spectrum. We assign
a probability to this point determined from how well it agrees with
our ``measured quantities'' using our estimates of the errors on those
quantities. The process is repeated for many points and the
probabilities used to determine the central values of the parameters,
their errors and their correlations. The probability distribution
functions are not always Gaussian; the $+1\sigma$ ($-1\sigma$) errors
quoted below are such that 15.87\% of 
this distribution is below (above) the quoted value.
Thus, 68.27\% of the
probability falls within our definition of $\pm 1\sigma$.

\subsection{LHC Point 3}

At LHC Point~3 ($m_0=200$ GeV, $m_{1/2}=100$ GeV), LEP will discover
$h$ and measure its mass. The relevant error is that from theoretical
calculations of the mass in the supergravity model which is likely to
dominate the error from the LEP measurement.  We will assume an error
of $\pm 3\,\GeV$.  Using the results presented in
Section~\ref{sec:lhc3}, we assume the following measurements and
errors:
\begin{itemize}
\item $M_{\tchi_2^0}-M_{\tchi_1^0}= 52.36 \pm 0.05$~GeV,
\item $M_{\tilde{g}}-M_{\tilde{b}}=20.3 \pm 2.0$~GeV,
\item $M_h=68.3 \pm 3$~GeV.
\end{itemize}
As described above the mass difference $M_{\tilde{g}}-M_{\tilde{b}}$ is
insensitive to the mass assumed for $\tchi_1^0$.  

Using the strategy outlined above we get the following constraint on
the parameters:  
\begin{itemize}
\item $m_0=200^{+13}_{-8}$~GeV, 
\item $m_{1/2}=99.9 \pm 0.7$~GeV,
\item $\tan\beta=1.99\pm 0.05$,
\item the sign of $\mu$ is determined to be $-1$,
\item $A_0$ is constrained to be greater than $-400$~GeV.  
\end{itemize}
There are no clear correlations between the parameters.  The
additional constraint that the average value of the light squark mass
is within $\pm 20$ GeV provides no additional restriction on the
parameters.  The relevant phenomenological parameter is the value of
$A$ evolved down to the the electroweak scale. The relevant values are
those for the third generation, $A_b$ and $A_t$.  Information on these
can only come directly from data on bottom and top squarks.  In this case
$A_t=-176 \pm 22\,\GeV$. $A_b$ is not well constrained; it is allowed
to range from $50$ to $-500$ GeV. Over this range the mass of the two
stop eigenstates varies only slightly from 270 and 320 GeV at one end
to 260 and 330 GeV at the other.  Constraining $A_b$ is very
challenging as it appears scaled by the small factor of the bottom quark Yukawa
coupling.

The degree of precision may be surprising. Over most of the SUGRA
parameter space $\tchi_2^0$ and $\lsp$ are gauginos ({\it i.e.}, they
have no Higgsino components), $M_{\tchi_2^0}-M_{\lsp}$ then determines
$m_{1/2}$. $M_{\tilde{g}}$ and $M_{\tchi_1^0}$ are then predicted and a
consistency check of the model made by the measurement of $M_{\tilde
g}$.  Information on $m_0$, is then obtained from $M_{\tilde{b}}$ and
the value of $M_h$ is sufficient to constrain $\tan\beta$.

\begin{figure}[h]
\dofig{3.20in}{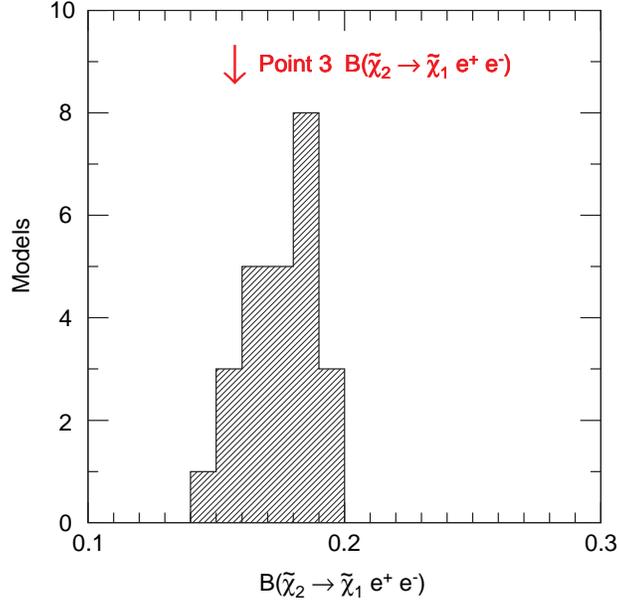}
\caption{Branching ratio $\tchi_2^0\to \tchi_1^0\ell^+\ell^-$ for the
SUGRA models $m_0=200 \pm 15 $ GeV, $m_{1/2}=100\pm 1.5$ GeV,
$\tan\beta=2.0 \pm 0.1$.\label{branchfig}}
\end{figure}

The other measurements available at this point are now used to provide
powerful consistency checks of the model.  The measurement of the
bottom squark and other squark masses and the event rates for isolated
leptons without jets discussed in Section~\ref{sec:lhc3} are examples.
Another example is the branching ratio for $\tchi_2^0\to
\tchi_1^0\ell^+\ell^-$. Figure~\ref{branchfig} shows this branching
fraction for a selection of SUGRA models that have parameters in the
slightly larger range $m_0=200 \pm 15 $ GeV, $m_{1/2}=100.\pm 1.5$ GeV,
$\tan\beta=2.0 \pm 0.1$.  At Point 3 $BR(\tchi_2^0\to
\tchi_1^- e^+e^-)=16.5\%$. This branching ratio can be constrained using the 
method described in section~\ref{sub:br};.

\subsection{LHC Point 5}
\label{sec:scan-5}

At LHC Point~5 ($m_0=100\,\GeV$, $\mhalf=300\,\GeV$, $\tan\beta=2.1$),
$h$ will be discovered at LHC in its decay to $b\bar{b}$ from its
production in the decays of supersymmetric particles, and its mass will
be measured precisely from its decay to $\gamma\gamma$.  
Using the results presented in Section~\ref{sec:lhc5}, we
assume the following set of measurements:
\begin{itemize}
\item$M_{\tchi_2} \sqrt{1-M_{\tilde{l}}^2/M_{\lsp}^2}
\sqrt{1-M_{\tchi_1^0}^2/M_{\tilde{l}}^2}= 108.6\pm 1$~GeV,
\item The decay $\tilde{g}\to \tilde{t} t$ is allowed,
\item The end point of the spectrum in Figure~\ref{c5_mbbj} is $506\pm
40$ GeV,
\item $M_h=104.15 \pm 3$ GeV.
\end{itemize}
These results correspond to two possible solutions:
\begin{itemize}
\item $m_0=100.5^{+12}_{-8}$ GeV,
\item $m_{1/2}=298^{+16}_{-9}$ GeV, 
\item $\tan\beta=1.8^{+0.3}_{-0.5}$, 
\item $\mu=+1$;
\end{itemize}
and
\begin{itemize}
\item $m_0=91 \pm 3 $ GeV, 
\item $m_{1/2}=288 \pm 18$ GeV, 
\item $\tan\beta=3.1 \pm 0.2$,
\item $\mu=-1$.
\end{itemize}
Both of these solutions provide good fits to the ``data''.  While
there is no constraint on $A$, the values of $A_t$ and $A_b$ are
constrained: $A_b=-740\pm 180 $ ($A_b=-750\pm 220$) GeV and
$A_t=-495\pm 30$ ($A_t=-536\pm 56$ ) GeV for $\mu$ positive (negative).
The main differences between the mass spectra for these two solutions
are in the masses of $\chi_2^0$, $\chi_3^0$, $\chi_4^0$, and
$\chi_2^+$ and the masses of the heavier Higgs bosons; they are
significantly larger in the positive $\mu$ case.  The claimed
sensitivity to the lepton decay spectrum (see
Figure~\ref{c5p20_pt2pt1}) of $\delta m_0\sim 5$ GeV implies that this
parameter's range can be narrowed somewhat. We have not investigated
the sensitivity of this spectrum to the sign of $\mu$.  At this point
the errors used (except for the one on $M_h$) are limited by
statistics, so additional luminosity will cause the errors to drop.
Using 0.6 GeV for the error on the dilepton endpoint and 23 GeV for
the error on the endpoint of Figure~\ref{c5_mbbj} reduces the errors
on $m_{1/2}$ and $m_0$ to $\pm 7$ and $\pm 9$ ($\pm 2.5$ and $\pm 10$)
respectively for positive (negative) $\mu$; the error on $\tan\beta$
is not reduced.

\begin{figure}[h]
\dofig{3.20in}{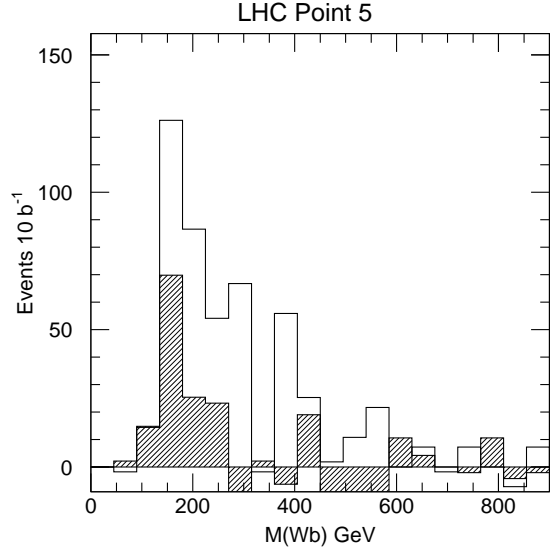}
\caption{The sideband subtracted invariant mass distribution of $Wb$
candidates. The analysis is as described in
section~\protect\ref{subsec:top} The solid histogram is for Point~5 and
is the same as that shown in Figure~\protect\ref{c5-marjietop} except
that the binning has been changed. The hatched histogram 
corresponds to the point $m_0=90.4\,\GeV$, $\mhalf=290\,\GeV$,
$\tan\beta=3.1$, $\sgn\mu=-1$ and $A_0=66$~GeV.
\label{c51-marjietop}}
\end{figure}

As well as the differences in the masses of the heavier Higgs and
gauginos noted above, the negative sign solution has larger stop
masses and hence a smaller branching ratio for $\tg \ra \tilde{t}_1
\bar{t}$. We have investigated the sensitivity of the top quark signal
discussed in section \ref{subsec:top} to the sign of $\mu$.  The
parameters $m_0=90.4\,\GeV$, $\mhalf=290\,\GeV$, $\tan\beta=3.1$,
$\mu=-1$, and $A=66$ GeV have the opposite sign of $\mu$ from Point~5
but give an excellent fit to the ``data'' used for fitting at this
point.\footnote{This modified point has a combined probability of 98\%
of fitting the data.} A sample of 100K events was generated for this
parameter set, and the analysis leading to Figure~\ref{c5-marjietop}
was repeated. The result is shown in Figure~\ref{c51-marjietop}; note
that the bins in this figure are twice as wide as those in
Figure~\ref{c5-marjietop}. It can be clearly seen that the amount of
reconstructed top in SUSY events is reduced relative to that at
Point~5 and that this fact can be used to eliminate this
alternative solution. In addition, the solutions with $\mu=-1$ have smaller branching ratios
of $\tilde{q}\to q \lsp h $ so the observed number of higgs events should be able to severely constrain this case.

\subsection{LHC Point 4}

At LHC Point~4 ($m_0=800\,\GeV$, $\mhalf=200\,\GeV$, $\tan\beta=10$),
determination of the parameters cannot be done by the simple method
already described. Here only two masses can be measured in a
straightforward manner; the light Higgs mass from its decay to
$\gamma\gamma$ and the $\tchi_1^0-\tchi_2^0$ mass difference from the
endpoint in the dilepton mass distribution.  We use these two
measurements and the determination of the SUSY scale from the
$\Meff$ analysis of Section~\ref{sec:meff}:
\begin{itemize}
\item $M_{\tchi_2^0}-M_{\tchi_1^0}= 69 \pm 1$,
\item $M_h=117.4 \pm 3$~GeV,
\item $\min(M_{\tg}, M_{\tilde u_R})=580 \pm 60$ GeV from the $\Meff$
analysis. 
\end{itemize}
These constraints restrict the parameter space to two regions:
\begin{itemize}
\item $m_0=784^{+203}_{-262}$~GeV,
\item $m_{1/2}=200 \pm 8$ GeV,
\item $\tan\beta=9 \pm 2$, 
\item $A$ is not constrained,
\item $\mu=+1$,
\item $A_b < 200 $ GeV and $A_t<  -150 $ GeV;
\end{itemize}
or
\begin{itemize}
\item $\tan\beta=14 \pm 4$, 
\item $m_0=950 \pm 210$,
\item $m_{1/2}=185 \pm 10$ GeV, 
\item $A$ is not constrained,
\item $\mu=-1$
\item $A_b=-160\pm 150 $ GeV and $A_t=-400\pm 100 $GeV.
\end{itemize}
The uncertainty on $M_{\tchi_2^0}-M_{\tchi_1^0}$ is limited by
statistics at low luminosity. A reduction in its error would reduce
the error on $m_{1/2}$.

In order to constrain parameters further, models whose parameters are
consistent with these values would need to be generated and their
predictions for the distributions shown in Figures~\ref{c4_mllm0},
\ref{c4_mllemu}, and \ref{c4_mllemudif} calculated. Those which are
inconsistent with the ``observed'' distribution can then be rejected.
This exercise is beyond the scope of this paper.

\subsection{LHC Points 1 and 2}

At LHC Point~1 and 2 ($m_0=400\,\GeV$, $\mhalf=400\,\GeV$,
$\tan\beta=2,10$), event rates are low and precision measurements
difficult at low luminosity.  We use the following constraints.
\begin{itemize}
\item $M_h=111.4 \pm 3$~GeV for Point~1, or 
\item $M_h=125.4 \pm 3$~GeV for Point~2, 
\item $\min(M_{\tg}, M_{\tilde u_R})=920 \pm 90$ GeV from the $\Meff$,
\item The end point of the spectrum in Figure~\ref{c5_mbbj} is $745\pm
50$~GeV.
\end{itemize}
In the case of Point~1, there are two solutions:
\begin{itemize}
\item $m_0$ unconstrained,
\item $m_{1/2}=400^{+40}_{-50}$ GeV,
\item $\tan\beta=2.0^{+0.4}_{-0.5}$, 
\item $\mu=+1$,
\item $A$ is not constrained,
\item $A_b=-1100\pm 200$ GeV and $A_t=-650\pm 55 $GeV;
\end{itemize}
and one with negative $\mu$ that has instead
\begin{itemize}
\item $m_{1/2} = 392^{+40}_{-50}$~GeV,
\item $\tan\beta=3.3 ^{+0.5}_{-0.4}$.  
\end{itemize}
In the case of Point~2, there is  again a solution for either sign of
$\mu$:
\begin{itemize}
\item $m_0$ unconstrained,
\item $m_{1/2}=405^{+32}_{-37}$~GeV,
\item $\tan\beta =10.6\pm 0.3$, 
\item $\mu=\pm 1$,
\item $A$ is not constrained,
\item $A_b=-1100\pm 200 $ GeV and $A_t=-800\pm 55 $GeV.
\end{itemize}
 
The lack of a constraint on $m_0$ at these points is alarming but can
be explained.  $m_{1/2}$ is large and the renormalization group
scaling from the GUT scale forces the squark masses to be comparable
to the gluino mass almost independent of the input value of $m_0$. In
cases of this type one needs to measure slepton masses which are less
affected since their renormalization group scaling is controlled by
$\alpha_{\rm weak}$ rather than $\alpha_s$. Since the slepton masses are
of order 500 GeV, this is a difficult task.  Even a lower bound on the
masses would constrain $m_0$.

If we reduce the error on the end point of the spectrum in
Figure~\ref{c5_mbbj} to 20 GeV which might be achievable with high
luminosity running at LHC, the uncertainty on $m_{1/2}$ reduces to
$\pm 30 $ GeV. As in the case of Point~5, the decay $\tg\to
\tilde{t}t$ is allowed. An analysis similar to that discussed there
should be able to establish that this channel is open.

\begin{figure}[h]
\dofig{3.20in}{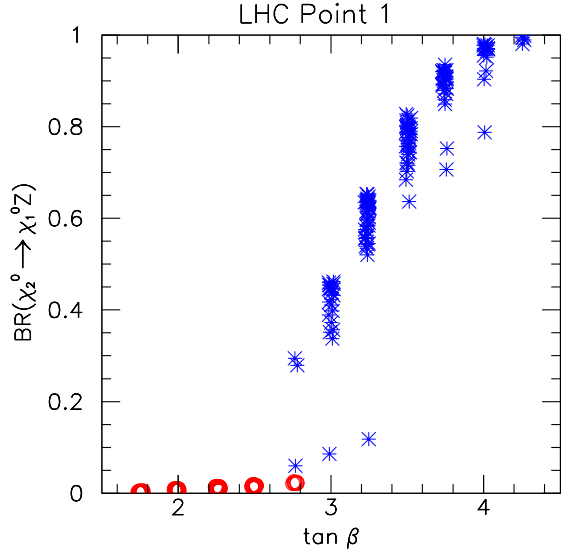}
\caption{The branching ratio $\chi_2^0\to \lsp Z$ for models with
$M_h=111.5 \pm$ 3 GeV, $\Meff=920 \pm 90$ GeV and both signs of $\mu=+1$ (circles)
and $\mu=-1$ (crosses) as a
function of $\tan\beta$.
\label{c1-branching}}
\end{figure}

In the case of these points, the negative $\mu$ solution can be eliminated.
The branching ratio for $\tilde{\chi_2^0} \to \lsp Z$
is shown in Figure~\ref{c1-branching} as a function of $\tan\beta$ for both
signs of $\mu$ for the solutions in the allowed range. It can be seen from this Figure that
the branching ratio is substantially larger for $\mu=-1$. The decay chain
$\tilde{q}_L\to q \tilde{\chi_2^0}\to 
q Z \lsp$ leads to a small $Z$ peak as shown
in Figure~\ref{c1_mll} which corresponds to
 $BR(\tilde{\chi_2^0}\to \lsp Z)=0.6\%$.
The smallest branching ratio for the $\mu=-1$ solution shown in Figure~\ref{c1-branching}
is 5.7\%, approximately a factor of ten larger. It is clear that these two cases can be 
distinguished in 10fb$^{-1}$ despite the fact thata larger data set may be required to measure the size of the peak
shown in Figure~\ref{c1_mll}. 
In the case of Point~2, 
The difference in branching ratios is even greater, again enabling elimination of the $\mu=-1$ solution.

\newpage

\newpage

\newpage
\section{Conclusions}

In this paper we have outlined a strategy that can be used to
systematically explore supersymmetry assuming that it is discovered at
the LHC.  We have given an example of a global variable ($\Meff$)
that can be used to determine the mass scale of SUSY if nature has
chosen the SUGRA model. Such global variables will be used to give the
first indication of a signal independent of the type of SUSY model.
The production of heavy mass states virtually guarantees that events
with very energetic jets will exist. Other SUSY models such as those
with $R-$parity violation \cite{hall} may have no missing $E_T$ if the
LSP decays within the detector. If the LSP decays to leptons, then all
SUSY events will have leptons and they can be used in a global
variable. If the LSP decays to hadrons, the jet multiplicity will rise
and a variable similar to $\Meff$ should be effective.

More detailed exploration will depend on the particular SUSY model.
In the SUGRA models, there is a general feature, namely that the
second lightest neutralino almost always decays to $h+\lsp$ if the
channel is open and to $\ell^+\ell^-+\lsp$ with a substantial
branching fraction if it is not.  In the former case, this will be the
dominant source of $h$ and it will be discovered in this process via
its decay to $b\overline{b}$ if it has not been seen at LEP.  In the
latter case the measurement of the position of the end point in the
$\ell^+\ell^-$ mass distribution provides a very precise measurement
of the mass difference between two of the sparticles. After first
observing one of these signals, one will move up the decay chain to
determine other quantities.

We have then illustrated, using specific examples, some techniques
that can be used to determine masses and branching ratios of
sparticles. Some of these quantities were then used to determine the
fundamental parameters of the SUGRA model some of which can be
determined with great precision. The ultimate goal of such studies
would be to use very many measurements to make an overconstrained fit
to the model, rather in the same way that current data are used to
test the Standard Model \cite{lankg}

The results in this paper are only an indication of the exciting
physics that lies ahead for the members of the CMS and ATLAS
collaborations if nature proves to be supersymmetric on the weak scale.

This work was begun during the Summer Study organized by the Division of
Particles and Fields of the American Physical Society. We benefited from many
discussions with attendees at that study and with colleagues on the 
ATLAS collaboration. In particular we would like to thank Stephan Lammel, John
Bagger, Alfred Bartl, Fabiola Gianotti and Daniel Froidevaux.

\bigskip

The work was supported in part by the Director, Office of Energy
Research, Office of High Energy Physics, Division of High Energy
Physics of the U.S. Department of Energy under Contracts
DE--AC03--76SF00098 and DE-AC02-76CH00016.  Accordingly, the U.S.
Government retains a nonexclusive, royalty-free license to publish or
reproduce the published form of this contribution, or allow others to
do so, for U.S. Government purposes. One of us (JS) would like to thank the
Swedish National Research council for support.


\end{document}